\documentclass[journal]{IEEEtran}
\usepackage{times} 
\usepackage{epsfig}
\usepackage{graphicx}
\usepackage{epstopdf}
\usepackage{algorithm,algorithmic,color,subfigure, latexsym, amsmath, amsfonts, amssymb, cite}
\usepackage{algorithm}
\usepackage{algorithmic}
\usepackage{epsfig}
\usepackage{graphicx}
\usepackage{epstopdf}
\usepackage{bm,amsmath,amssymb,amsfonts,graphicx,epsfig,amsthm,color}
\usepackage{bbold,dsfont}
\usepackage{float}
\usepackage{hhline}
\usepackage[hyphens]{url}
\PassOptionsToPackage{bookmarks=false}{hyperref}

\usepackage[depth=-1]{bookmark}

\usepackage{booktabs}       
\usepackage{amsfonts}       
\usepackage{microtype}      

\usepackage{times}
\usepackage{graphicx} 

\usepackage{wrapfig}



\usepackage{accents}
\makeatletter
\def\wid{\check{{\cc@style\underline{\mskip9.5mu}}}}
\def\Wideubar{\underaccent{{\cc@style\underline{\mskip6mu}}}}
\makeatother

\makeatletter
\def\wideubar{\underaccent{{\cc@style\underline{\mskip9.5mu}}}}
\def\Wideubar{\underaccent{{\cc@style\underline{\mskip6mu}}}}
\makeatother

\makeatletter
\def\widebar{\accentset{{\cc@style\underline{\mskip9.5mu}}}}
\def\Widebar{\accentset{{\cc@style\underline{\mskip6mu}}}}
\makeatother

\newtheorem{Lemma}{Lemma}

\newtheorem{Theorem}{Theorem}

\newtheorem{Assumption}{Assumption}

\theoremstyle{Remark}\newtheorem{Remark}{Remark}

\allowdisplaybreaks
\begin{document}
	
	\title{\Large \bf Data-Driven Control of {Distributed} Event-Triggered Network Systems}
	\author{
		Xin Wang,~Jian Sun,
		Gang Wang, 
		Frank Allg{\"o}wer,
		and~Jie Chen
		\thanks{This work was supported in part by the National Key R$\&$D Program of China under Grant 2021YFB1714800, the National Natural Science Foundation of China under Grants 62088101, 61925303, 62173034, U20B2073, and 61720106011, and in part by the Natural Science Foundation of
Chongqing under Grant 2021ZX4100027.}
		
		\thanks{
			X. Wang, J. Sun, and G. Wang are with the State Key Lab of Intelligent Control and Decision of Complex Systems, School of Automation, Beijing Institute of Technology, Beijing 100081, China, and also with the Beijing Institute of Technology Chongqing Innovation Center, Chongqing 401120, China (e-mail: xinwang@bit.edu.cn, sunjian@bit.edu.cn, gangwang@bit.edu.cn).

			F. Allg{\"o}wer is with the Institute for Systems Theory and Automatic Control, University of Stuttgart, 70550 Stuttgart, Germany (e-mail: frank.allgower@ist.uni-stuttgart.de).
			
			J. Chen is with the Department of Control Science and Engineering, Tongji University, Shanghai 201804, China, and also with the State Key Lab of Intelligent Control and Decision of Complex Systems, School of Automation, Beijing Institute of Technology, Beijing 100081, China 	
			(e-mail: chenjie@bit.edu.cn).}
	}
	\maketitle
	
	\begin{abstract}
		The present paper deals with data-driven event-triggered control of a class of unknown discrete-time interconnected systems (a.k.a. network systems). To this end, we start by putting forth a novel distributed event-triggering transmission strategy based on periodic sampling, under which a model-based stability criterion for the closed-loop network system is derived, by leveraging a discrete-time looped-functional approach. Marrying the model-based criterion with a data-driven system representation recently developed in the literature, a purely data-driven stability criterion expressed in the form of linear matrix inequalities (LMIs) is established. Meanwhile, the data-driven stability criterion suggests a means for co-designing the event-triggering coefficient matrix and the feedback control gain matrix using only some offline collected state-input data. Finally, numerical results corroborate the efficacy of the proposed distributed data-driven ETS in cutting off data transmissions and the co-design procedure.
	\end{abstract}
	
	
	
	\section{Introduction}
	
	Network systems have gained enormous attention over the last two decades, thanks to their widespread applications in diverse science and engineering disciplines \cite{Dimarogonas2012,JAS2022lhy,Bansal2020distributed,eng2022,Ding2021Review,Wang2022delay}.
	Such systems are usually composed of a group of interconnected subsystems distributed across large geographical locations, and information is exchanged through a public communication network. However, limited network resources, e.g., bandwidth and  energy of wireless transmitting nodes, make the control design challenging \cite{cdc2012heemels,Ge2021}.
	Traditional sampled-data schemes, which are also known as time-triggered control, feature low computational overhead and easy deployment,
	but they often incur a considerable number of ``redundant" transmissions occupying network resources \cite{Donkers2012,Chen2020,Chen2021event,Wang2021mixed}.
	
	Recently, for network systems, research efforts have centered around devising event-triggering transmission strategy that can rely on minimal communication resources while maintaining acceptable control performance. 
	A decentralized ETS (ETS) using only local information was proposed for distributed control of continuous-time network systems in \cite{Wangxf2011,Wang2015event}, which was extended to discrete-time network systems in \cite{SHI2019}.
	Such decentralized ETSs are not optimal for distributed systems,
	because they did not exploit the interactions between adjacent subsystems in the design. By adding the neighbors' information, a distributed ETS for multi-agent linear systems was designed in \cite{Dimarogonas2012}. 
	In \cite{Deng2021}, a dynamic periodic distributed ETS was designed by introducing a dynamic variable into the static periodic distributed ETS \cite{GUO2014}, which can effectively reduce transmissions for each agent.
In the context of networked control, continuous-time systems are controlled via digital computers, in which case one typically first discretizes the continuous-time system and works with the resulting discrete-time counterpart.
As a matter of fact, few results on discrete-time ETSs for network systems are available in the literature.

	
	
	
	%
	
	On the other hand, all the above-mentioned ETSs are model-based, in the sense that they require explicit knowledge of system models for the controller design and implementation.
	Nevertheless, obtaining an accurate model of a real-world system can be computationally expensive and even impossible in many practical engineering situations. Furthermore, the obtained models are just too complex for classic control methods to be employed \cite{Astrom1973article}.
	An alternative approach, which is termed \emph{data-driven control}, has recently attracted considerable attention. Data-driven control is aimed at performing control laws directly from measured data without requiring any steps of identifying real systems.
	Under this umbrella, various results  have been presented, see summaries in e.g., \cite{Hou2013,markovskyabehavioral2021,ren2021tii,Jaap2022Informativity}.  
	By joining the data-based system parametrization in \cite{Waarde2020} and the model-based analysis method in \cite{Fridman2010}, a data-driven stability criterion was developed for \emph{continuous-time} linear systems with sampled-data control in \cite{Berberich2020}, which has also suggested a data-driven controller. This data-driven framework has been extended to \emph{discrete-time} sampled-data systems in \cite{wildhagen2021datadriven}. 
	A data-driven resilient predictive controller against DoS attacks was developed in \cite{Liu2021data}.
	Data-driven control of event-triggered continuous-time systems with constant delays was recently investigated in \cite{Wang2021data}, which was then extended to the case of the discrete-time domain by \cite{wang2022disdata}.
	Thus far, it remains an unexplored territory to co-design data-driven distributed controllers as well as event-triggering transmission schemes for unknown discrete-time networked interconnected systems.
	
	These developments have motivated our research on data-driven control of discrete-time networked systems under distributed event-triggering transmissions in this paper.
	In existing distributed ETSs \cite{GUO2014,Deng2021}, only the relative state information of subsystem's neighbors was used to design the ETS.
{The subsystem state used in designing the threshold of decentralized ETSs \cite{Wangxf2011,SHI2019} was not considered in \cite{GUO2014,Deng2021}.
	An undesired low triggering frequency may occur, when only the state errors between adjacent subsystems are considered and they are less than the local subsystem state, therefore considerably deteriorating the system performance.}
	
	In this submission, we develop a new dynamic distributed event-triggering transmission strategy based on periodic sampling for discrete-time network systems, {which generalizes the aforementioned decentralized and distributed ETSs.} 
	In addition, in terms of controller design and stability analysis, looped-functionals \cite{SEURET2012177} have recently been shown competitive relative to common Lyapunov functionals that were used in existing distributed event-triggered control.
	By virtue of a discrete-time looped-functional, we derive a model-based criterion for distributed event-triggered control of network systems.
	By combining the model-based criterion and the data-driven representation of discrete-time linear systems developed in
	\cite{Waarde2020},
	a data-based stability criterion is established, which provides a data-driven method for co-designing the state feedback control gain matrix as well as the matrix in the event-triggering condition.
	
	In succinct form, the  contribution of this paper is two-fold, summarized as follows.
	\begin{enumerate}
		\item [\textbf{c1)}] We develop a novel dynamic distributed ETS on the basis of periodic sampling for discrete-time {network systems, where the triggering law only involves the discrete-time state information of each subsystem and its neighbor(s) at sampling instants};
		\item [\textbf{c2)}] We establish a model-based stability criterion for discrete-time network systems under the proposed distributed ETS using a tailored discrete-time looped-functional, along with model- and data-driven co-designing approaches of the controller and the ETS.
	\end{enumerate}
	
	{The remainder of the paper is organized as follows. In Section
\ref{Sec:preliminaries}, we introduce the network system under a distributed sample-data control strategy, as well as the data-based system
representation. Section \ref{sec:eventtrigger} provides the main results including model- and data-driven design methods of
the controller for the network system with a novel distributed transmission scheme.
Section \ref{sec:example} validates the merits and effectiveness of our methods by numerical simulations. Section \ref{sec:conclusion} summarizes this paper.}

	{\it Notation.}
	Throughout the paper, let symbols $\mathbb{R}^n$, $\mathbb{R}^{n\times m}$, and $\mathbb{N}$ denote accordingly the set of all $n$-dimensional real vectors, ${n\times m}$ real matrices, and non-negative
	integers, respectively. For any integers $a,b\in \mathbb{N}$,  define 
	$\mathbb{N}_{[a,b]}:=\mathbb{N}\cap [a,b]$.
	Furthermore, we write $P\succ 0$ ($P\succeq 0$) if $P$ is a symmetric positive (semi)definite matrix, and $[\cdot]$ if its elements can be inferred by symmetry. Symbol ${\rm blkdiag}\{\cdots\}$ represents block-diagonal matrices, `$\ast$' the symmetric term in (block) symmetric matrices, and
	${\rm Sym}\{P\}$ the sum of $P^{T}$ and $P$. Finally,
we use $0$ ($I$)  to stand for zero (identity) matrices
	of appropriate
	dimensions.

	
	\section{Problem Formulation}\label{Sec:preliminaries}
	\subsection{Sampled-data control of network systems}

Consider a network system consisting of $N$ discrete-time linear time-invariant subsystems (a.k.a. agents). These subsystems are coupled together and controlled by local controllers. A communication network is available to and shared by distributed systems.  Each agent can communicate with its neighbors through the network, while the local controller can receive the neighbors' information.
The information flow can be described by an undirected graph
$\mathcal{G}:=\{\mathcal{V},\mathcal{E},\mathcal{C}\}$, where $\mathcal{V}:=\{v_1, v_2, \ldots,v_N\}$ is the set of nodes, and $\mathcal{E} \subseteq \mathcal{V}\times \mathcal{V}$ represents the set of edges. The matrix $\mathcal{C}:=[c_{ij}]\in \mathbb{R}^{N\times N}$ is the adjacency of $\mathcal{G}$, constructed by setting $c_{ij}=c_{ji}=1$ if node $v_i$ and node $v_j$ are adjacent and they can share information between each other via communication channels and $c_{ij}=c_{ji}=0$, otherwise.  Self-loops
are not taken into consideration, i.e., $c_{ii}=0$ for all $i\in\mathbb{N}_{[1,N]}$.

\begin{figure}
		\centering
		 \includegraphics[scale=0.5]{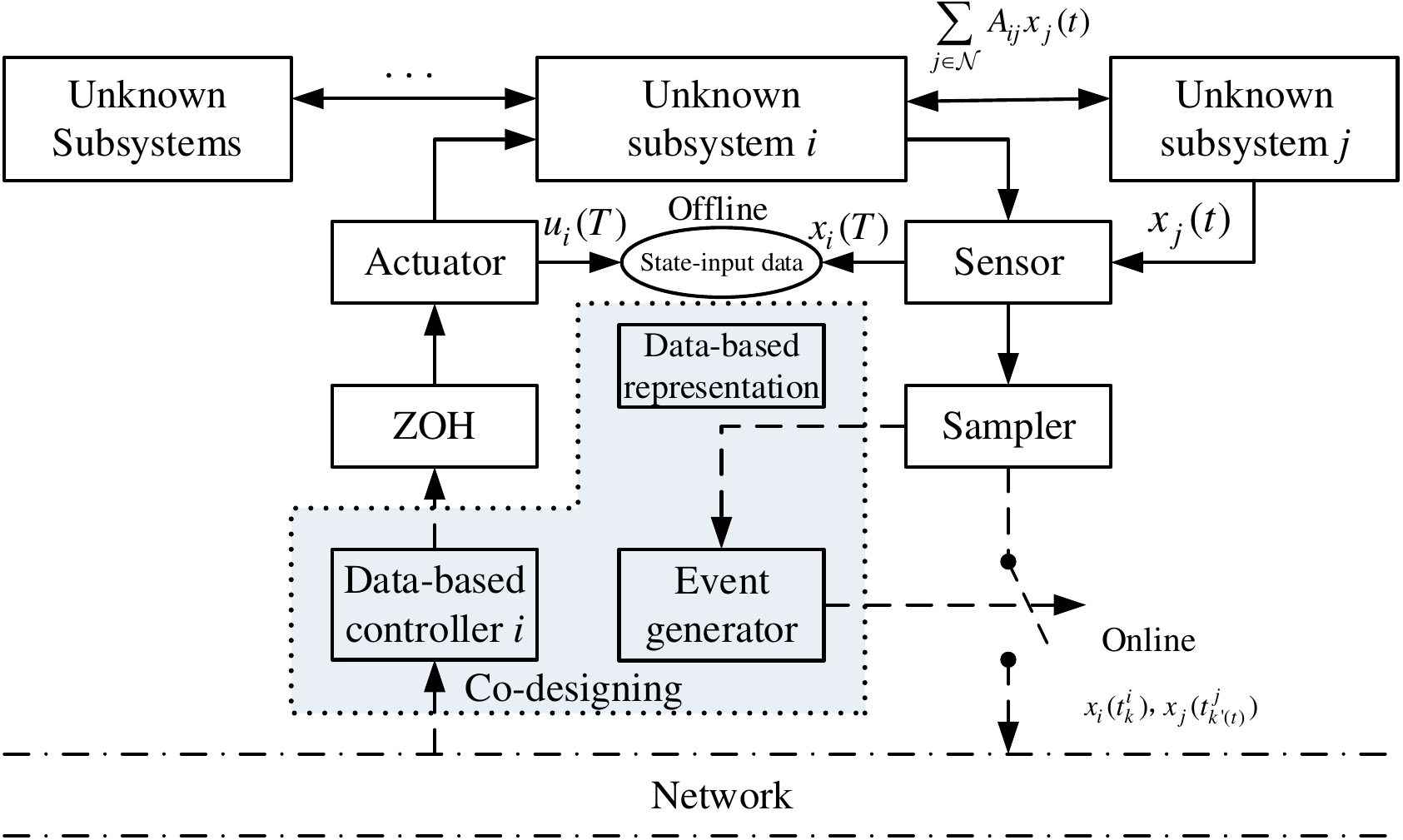}
		\caption{Data-driven networked control under ETS.}
		\label{FIG:structure:event}
	\end{figure}

The state equation of subsystem $i$ is given by
	\begin{equation}\label{sec1:sys:disLTI}
		x_i(t+1)=A_i x_i(t)+\sum_{j \in \mathcal{N}} A_{ij}x_j(t)+B_i u_i(t), ~t\in \mathbb{N}
	\end{equation}
	for all $i\in \mathbb{N}_{[1,N]}$ and $\mathcal{N}:=\{j\in \mathbb{N}_{[1,N]}|j\neq i, c_{ij}=1\}$, where $x_i(t)\in \mathbb{R}^{n_i}$ and $u_i(t)\in \mathbb{R}^{m_i}$ are the state vector and the control input of subsystem $i$, respectively; {$t\in \mathbb{N}$ denotes the discrete time index}; and,  $A_i\in\mathbb{R}^{n_i\times n_i}$, $B_i\in\mathbb{R}^{n_i\times m_i}$, $A_{ij}\in\mathbb{R}^{n_i\times n_j}$ are its system matrices.
Specifically, if the subsystem $i$ can physically interact with the subsystem $j$, it holds that $A_{ij}\neq 0$; and $A_{ij}= 0$ when there is no interconnection between the subsystems $i$ and $j$.
See Fig. \ref{FIG:structure:event} for an illustration, where each subsystem can communicate with its neighbor(s), and the communication topology is captured by the matrices $\{A_{ij}\}$. 


	In fact, such network systems can model a wide range of applications, and they have been largely investigated in the literature, see e.g., \cite{Wangxf2011},
	\cite{auto2019nowzari}, \cite{GUAN2020,JAS2022lhy}.
	In contrast to the existing works, this paper focuses on a more challenging situation, in which the system matrices $A_i$, $B_i$, and $A_{ij}$ are all assumed \emph{unknown}.
	To save computation and communication resources, event-triggered transmissions between neighboring subsystems are considered. The overall goal of this paper is to develop purely data-driven solutions for co-designing the stabilizing control laws as well as the event-triggering protocols for all subsystems without explicit knowledge of the system matrices $A_i$, $B_i$, $A_{ij}$.
	
	It is assumed that during the closed-loop operation, the state of each subsystem is periodically sampled (e.g., with a common period) in a synchronous manner, and the sampled state is transmitted to local and neighbors' controllers only at event-triggered time instants. The event times of agent $i$ are denoted by  $t_0^i,~t_1^i,\,\ldots,t_k^i,\ldots\,$, {where $k \in \mathbb{N}$ represents the order sequence of the transmission times} and
	\begin{equation}\label{sys:instants}
		t_0^i=0,~~ t_{k+1}^i-t_k^i\geq1.
	\end{equation}
In distributed control, each agent updates its own control input at event times it dictates by capitalizing on all information locally available as well as received from its neighboring agents.
	Specifically, at agent $i$, the local sampled state $x_i(t_k^i)$ and the states $x_j(t_{k'(t)}^j)$ of the neighboring agents are available, and the control input is computed using the linear feedback law
	\begin{equation}\label{feedbackC}
		u_i(t)=K_ix_i(t_k^i)+\sum_{j \in \mathcal{N}} K_{ij}x_j(t_{k'(t)}^j),~~t\in \mathbb{N}_{[t_k^i, t_{k+1}^i-1]}
	\end{equation}
	where {$k'(t):={\rm arg} \min_{l\in \mathbb{N}:t\geq t_l^\varsigma} \{t-t_l^\varsigma\}$ for agent $\varsigma \in \mathbb{N}_{[1,N]}$,}
	 and therefore, for each $t\in \mathbb{N}_{[t_k^i, t_{k+1}^i-1]}$, $t_{k'(t)}^j$ is the last event time of agent $j$;
	  the controller gain matrices $K_i\in\mathbb{R}^{n_i\times n_i}$ and $K_{ij}\in\mathbb{R}^{n_i\times n_j}$ are to be
	   designed such that all subsystems are stabilized.
A nonzero coupling matrix $K_{ij}$ implies that there exists a communication channel through which controller $i$ can utilize
the state $x_j(t_{k'(t)}^j)$, otherwise $K_{ij}=0$.
	
	Upon collecting the states of all subsystems to form ${x_k(t)}:=[x_1^\top (t_{k'(t)}^1)~\cdots~ x_i^\top(t_{k}^i)~\cdots~x_N^\top (t_{k'(t)}^N)]^\top $,
	the state equation of the entire network system under the above-mentioned linear feedback controller can be collectively expressed as follows
	\begin{equation}\label{sys:sampling}
		x(t+1)=A x(t)+BK{x_k(t)},~t\in \mathbb{N}_{[t_k^i, t_{k+1}^i-1]}
	\end{equation}
	where $B:={\rm blkdiag}\{B_1~B_2~\cdots~B_N\}$, and
	\begin{align}
		A&:=\left[ \,\begin{array}{cccc}
			A_1&A_{12}&\cdots&A_{1N} \\
			A_{21}&A_2&\cdots&A_{2N}  \\
			\vdots & \vdots&\ddots&\vdots \\
			A_{N1}&A_{N2}&\cdots&A_N \\
		\end{array} \right]\\
		K&:=\left[ \begin{array}{cccc}
			K_1 &   K_{12} & \cdots & K_{1N} \\
			K_{21} & K_2 & \cdots & K_{2N}  \\
			\vdots  &  \vdots & \ddots & \vdots \\
			K_{N1} & K_{N2} & \cdots & K_N \\
		\end{array} \right].
	\end{align}

{
	\begin{Remark}[Distributed controller]\label{Remark:controller}
\emph{The proposed distributed controller in \eqref{feedbackC} has access to both the information of local agent and its neighbor(s) through the communication network. It has been discussed in  \cite{GUAN2020} that traditional decentralized control strategy \eqref{feedbackC} with $K_{ij}=0$  (using only local  sampled states) incurs reduced
design complexity and computational overhead. However, no exchange of information between adjacent subsystems may result in  performance degradation and limit  applications of the decentralized controller in network systems that contains large numbers of interactive nodes.
From the perspective of system trajectories, the proposed
distributed control strategy can achieve better performance than the decentralized one, as certificated by our numerical example in Section \ref{sec:compare:controller}.}
\end{Remark}}

	In the remainder of this paper, we discuss how to design stabilizing feedback controller gain matrices along with dynamic event-triggering strategies for system \eqref{sys:sampling} or
the interconnected linear systems  \eqref{sec1:sys:disLTI}, in a distributed and data-driven manner. We further develop theoretical analyses and performance guarantees for the resultant data-driven control solutions.
	
	\subsection{Data-driven system representation for noisy data}\label{subsection:data}
	
	A key challenge in this paper is to analyze system stability and design the controller without using the system matrices $A_i$, $B_i$, and $A_{ij}$.
	For this purpose, the data-driven parametrization of the linear discrete-time systems in \cite{Waarde2020} is recollected.
	Suppose we have locally collected for each subsystem data $\{x_i(T)\}^{\rho}_{T=0}$ and $\{u_i(T)\}^{\rho-1}_{T=0}$ $(T\in \mathbb{N},~\rho\in \mathbb{N}_{[1,\infty]})$, at discrete time instants $T \in \{0,1,\ldots,\rho\}$, from a perturbed version of system \eqref{sec1:sys:disLTI} given by
	\begin{equation}\label{sys:data:perturbed}
x_i(T+1)=A_i x_i(T)+\sum_{j \in \mathcal{N}} A_{ij}x_j(T)+B_i u_i(T)+B_{w_i}w_i(T)
	\end{equation}
 where $B_{w_i}\in \mathbb{R}^{n_i\times n_{w_i}}$ is a known matrix and is assumed to have full column rank, which can model the influence of disturbance on the subsystem.
	Here,
	the collected data are corrupted by an {\it unknown} disturbance sequence $\{w_i(T)\}^{\rho-1}_{t=0}$, where $w_i(T)\in \mathbb{R}^{n_w}$ captures, e.g., the noise or unmodeled system dynamics.
	Each local set of pre-collected data can be stacked up to form the following matrices for $i\in \mathbb{N}_{[1,N]}$ 
	\begin{align*}
		X_i^+ &:=\big[\begin{array}{cccc}x_i(1)&x_i(2)&\cdots &x_i(\rho) \\\end{array}\big]\\
		 X_i&:=\left[\begin{array}{cccc}x_i(0)&x_i(1)&\cdots &x_i(\rho-1) \\\end{array}\right]\\
		 U_i&:=\left[\begin{array}{cccc}u_i(0)&u_i(1)&\cdots &u_i(\rho-1) \\\end{array}\right]\\
		 W_i&:=\left[\begin{array}{cccc}w_i(0)&w_i(1)&\cdots &w_i(\rho-1) \\\end{array}\right]
	\end{align*}
	where the data matrices $X_i^+$, $X_i$, and $U_i$
	 are known, but the noise matrix $W_i$ is unknown.
Then,
	it is now obvious that one can write
	\begin{align}\label{formu:data:local}
		X_i^+=A_iX_i+\sum_{j \in \mathcal{N}} A_{ij}X_j+B_iU_i+B_{w_i}W_i.
	\end{align}
Through summarizing the local data, similar to \eqref{sys:sampling},
the equation corresponding to global data collection is given as follows
\begin{align}\label{formu:data}
		X^+=AX+BU+B_{w}W
	\end{align}
where $B_{w}={\rm blkdiag}\{B_{w_1}~B_{w_2}~\cdots~B_{w_N}\}$, and
\begin{align*}
X^+&:=\big[{X_1^+}^\top ~ {X_2^+}^\top~\cdots~{X_N^+}^\top \big]^\top \\
X&:=\big[X_1^\top ~~~X_2^\top~~\cdots~~X_N^\top\big]^\top \\
U&:=\big[U_1^\top ~~~U_2^\top~~\cdots~~U_N^\top\big]^\top\\
W&:=\big[W_1^\top ~~~W_2^\top~~\cdots~~W_N^\top\big]^\top.
\end{align*}

	In practice, the noise is always bounded.
It is thus reasonable to make the following assumption on the noise,
which has also been used in e.g., \cite{Berberich2020,wildhagen2021datadriven}. 
	\begin{Assumption}[Global noise bound]\label{Ass:disturbance}
		The noise sequence $\{w_i(T)\}^{\rho-1}_{t=0}$ $(i\in \mathbb{N}_{[1,N]})$ collected in the matrix $W$ belongs to the set
		\begin{align*}
			\mathcal{W}=\!\bigg\{W\in\mathbb{R}^{ Nn_w\times\rho} \Big |
			\left[ \begin{array}{cc}W^\top \\I \\\end{array}\! \right]^\top
			\left[ \begin{array}{cc}Q_d  &  S_d\\\ast  &  R_d \\\end{array}\right]
			\left[ \begin{array}{cc}W^\top \\I \\\end{array}\!  \right]\succeq0 \bigg\}
		\end{align*}
		where $S_d \in \mathbb{R}^{\rho\times Nn_w}$, $Q_d \prec 0 \in \mathbb{R}^{\rho\times \rho}$, and $R_d={R_d}^\top  \in \mathbb{R}^{ Nn_w\times  Nn_w}$ are known matrices.
	\end{Assumption}

Assumption \ref{Ass:disturbance} is a general form for modeling global bounded additive noise, which summarizes all local noises. When $N=1$, Assumption \ref{Ass:disturbance} is equivalent to the one used for centralized control systems by \cite{persis2020,Berberich2020,Waarde2020}.

	Based on Equation \eqref{formu:data} and Assumption \ref{Ass:disturbance}, we may define the set of all matrices $[A~B]$ that are consistent with the measurements and the bounded noise as
	\begin{align*}
		\Sigma_{AB}:=\Big\{[A~ B] \left | {X^+}=AX+BU+B_wW,~ W\in \mathcal{W}\right.\Big\}.
	\end{align*}

Recalling \cite[Lemma 4]{Waarde2020}, it can be shown that $\Sigma_{AB}$ can be equivalently expressed as a quadratic matrix inequality (QMI) detailed as follows. 

	\begin{Lemma}[Data-driven system representation] \label{Lemma:system:data}
		The set $\Sigma_{AB}$ is equivalent to
		\begin{align*}
			 \Sigma_{AB}=\bigg\{[A~B]\in\mathbb{R}^{n\times (n+m)} \Big |
			\left[\begin{array}{cc}[A~B]^\top \\I \\\end{array}\right]^\top
			\Theta_{AB}
			\left[\cdot\right]\succeq0
			\bigg\}
		\end{align*}
		where  $\Theta_{AB}:=
		\left[\begin{array}{cc}-X & 0 \\ -U & 0 \\ \hline {X^+} & B_w \\\end{array}\right]
		\left[\begin{array}{cc}Q_d & S_d\\\ast & R_d \\\end{array}\right]
		[\cdot]^\top .$
	\end{Lemma}

	Lemma \ref{Lemma:system:data} provides a purely data-driven parametrization of the \emph{unknown} system \eqref{sec1:sys:disLTI} using only data $X$, ${X^+}$, and $U$.
{	 Note that such data are collected from the perturbed system \eqref{sys:data:perturbed}, while we analyze the stability of the unperturbed sampled-data system \eqref{sys:sampling}.
Introducing a disturbance in the presentation is to account for possible noise in pre-collected state-input data, rather than accommodating an explicit
disturbance on the system. In order to
guarantee the stability of \eqref{sys:sampling} without relying on any
knowledge of the system matrices, we need to achieve a
stability criterion for all $[A~B]\in \Sigma_{AB}$.
Note also that, in Fig. \ref{FIG:structure:event}, such data are collected offline in
an open-loop experiment. In comparison, the sampling times
$\{t_k\}_k$ come from a single closed-loop operation, which are independent of the open-loop data sampling.}
	In the following section, we show the main result that provides a data-driven method for co-designing the distributed controllers and the ETS, where Lemma \ref{Lemma:system:data} is employed to describe the system matrices consistent with the data collected from \eqref{sys:data:perturbed}.

	\section{Main Results}\label{sec:eventtrigger}
	In this section, we delineate our distributed dynamic event-triggering strategy for discrete-time network systems in Subsection \ref{subsec:event}.  Model-based stability analysis of the distributed network system under the ETS is performed in Subsection \ref{sec:modle:stability}, while model- and data-driven co-design strategies of the distributed control and ETSs are studied in Subsections \ref{design:event:model} and \ref{design:event}, respectively.
	
	\subsection{Distributed dynamic ETS }\label{subsec:event}
	We implement a discrete-time distributed event-triggering module to dictate the transmission times $\{t_k^i\}_{k\in \mathbb{N}}$ for each subsystem $i$, depicted in Fig. \ref{FIG:structure:event}.
In such event-triggered network systems, it is assumed that all subsystems are sampled
at time instants $\{t_k^i+vh\}_{v,k,h \in \mathbb{N}}$, where $h$ is the sampling interval satisfying $1\leq \underline{h} \leq h \leq \bar{h}$ with given lower and upper bounds $\underline{h}$, $\bar{h} \in \mathbb{N}$.
{It should be mentioned that the sampling interval $h$ is not completely consistent with the step size of the discrete-time system \eqref{sec1:sys:disLTI}; that is, $h>1$ can be designed. Therefore, the trigger module is not required to detect the system state at every discrete time $t$.}
	Building upon the distributed periodic ETS in \cite{GUO2014,Deng2021,GUAN2020}, an event generator is introduced to determine whether the sampled data at $t_k^i+vh$ needs to be transmitted or not for each agent, capitalizing on the following criterion
	\begin{align}\label{sys:judgement}
		&\eta_i(\tau_v^i)+\theta_i \rho_i(\tau_v^i)<0,\quad t\in \mathbb{N}_{[t_k^i, t_{k+1}^i-1]}
	\end{align}
	where {$\theta_i>0$ is to be designed}; $\tau_v^i:=t_k^i+vh$ for all $v\in \mathbb{N}_{[0,m^i_k]}$ with  $m^i_k=\frac{t_{k+1}^i-t_k^i}{h}-1$; and,
	\begin{align*}
		\rho_i(\tau_v^i):=&~\sigma_1^i x_i^\top (\tau_v^i)\Omega_i x_i(\tau_v^i)-e_i^\top (\tau_v^i)\Omega_i e_i(\tau_v^i)\\
		&+{\sum_{j\neq i}^N \sigma_2^{ij} \left[x_i(\tau_v^i)-x_j(\tau_v^i)\right]^\top \Omega_i \left[x_i(\tau_v^i)-x_j(\tau_v^i)\right]}
	\end{align*}
	where $\Omega_i\succ0$ is a weight matrix, and $\sigma_1^i$, $\{\sigma_2^{ij}\}_{j\in\mathcal{N}}$, $\theta_i$ are parameters, both to be designed ($\sigma_2^{ij}=0$ when there is no connection between the subsystems $i$ and $j$); $e_i(\tau_v^i):=x_i(\tau_v^i)-x_i(t_k^i)$ denotes the error of agent $i$ between sampled signals $x_i(t_k^i)$ at the latest sending
	instant and $x_i(\tau_v^i)$ at the current sampling instant; and, {$x_i(\tau_v^i)-x_j(\tau_v^i)$} 
represents the relative state information of agent $i$'s neighbors, which may enlarge the triggering intervals when compared with \cite{GUAN2020} that ignored the relative information.
	$\eta_i(\tau_v^i)$ in the condition \eqref{sys:judgement} is a discrete-time dynamic parameter satisfying the following difference equation
	\begin{align}\label{sys:dynamic}
		 \eta_i(\tau_{v+1}^{i})-\eta_i(\tau_v^{i})=-\lambda_i\eta_i(\tau_v^{i})+\rho_i(\tau_v^{i})
	\end{align}
	where $\eta_i(0)\geq0$ and $\lambda_i>0$ are given parameters. Different with the continuous-time one in \cite{Deng2021}, the extra dynamic variable $\eta_i(\tau_v^i)$ only changes at discrete sampling points $\tau_v^i$, which  eliminates the computational burden caused by continuously computing $\eta_i(t)$. Besides, it was shown in
 \cite{Girard2015} that such a dynamic scheme can further reduce the triggering frequency compared with the static strategy \cite{GUO2014}.

	If the condition \eqref{sys:judgement} is satisfied, the current local sampled state $x_i(\tau_v^{i})$ and the adjacent one $x_j(\tau_v^{i})$ are transmitted to the controller $i$, and a new control input is computed which is held via a zero-order holder (ZOH) in
	the interval $[t_{k+1}^i, t_{k+2}^i-1]$. Then, the trigger module is updated and waiting to trigger the next sampling event.
	In summary, our triggering law can be described as
	\begin{align}\label{sys:trigger}
		&t_{k+1}^i=t_k^i+h\cdot\min_{v\in \mathbb{N}}\Big\{v>0\Big|\eta_i(\tau_v^{i})+\theta_i\rho_i(\tau_v^{i}) < 0\Big\}.
	\end{align}

{The following lemma proves that the extra dynamic variable $\eta_i(\tau_v^{i})$ keeps positive for all $\tau_v^{i}$ if the initialization of the variable is greater than zero (cf. $\eta_i(0)\geq0$), when $\eta_i(\tau_v^{i})$ satisfies the  difference equation in \eqref{sys:dynamic}.} 

	\begin{Lemma}[\emph {Non-negativity.}]\label{lemma:nonneg.dynam}
		Let $\eta_i(0)\ge0$ be non-negative scalars, $\Omega_i\succ 0$ be positive semi-definite matrices, and $\lambda_i>0$, {$\theta_i> 0$} be constants satisfying $1-\lambda_i-\frac{1}{\theta_i}\geq0$.
		Then, for all $v\in \mathbb{N}_{[0,m^i_k]}$, it holds that $\eta_i(\tau_v^{i})\geq0$.
	\end{Lemma}
	
		\begin{IEEEproof}
	Executing the event-trigger in \eqref{sys:trigger}, the inequality \eqref{sys:judgement} is not satisfied for all $v\in \mathbb{N}_{[0,m^i_k]}$, which ensures that
	\begin{align}\label{Lemma1:1}
		\eta_i(\tau_v^{i})+\theta_i \rho_i(\tau_v^{i}) \geq0.
	\end{align}
	Inequality \eqref{Lemma1:1} suggests that $\rho_i(t) \geq -\frac{1}{\theta_i}\eta_i(\tau_v^{i})$.
	From \eqref{sys:dynamic}, it is obtained that
	\begin{align}\label{Lemma1:3}
		 \eta_i(\tau_{v+1}^{i})-\eta_i(\tau_{v}^{i})\geq -\lambda_i\eta(\tau_{v}^{i})-\frac{1}{\theta_i}\eta_i(\tau_{v}^{i}).
	\end{align}
	By induction, using the initial condition $\eta_i(0)\geq0$ and the assumption that {parameters $\lambda_i>0$ and $\theta_i>0$ satisfy}
$1-\lambda_i-\frac{1}{\theta_i}\geq0$, it follows that $\eta_i(\tau_{v}^{i})\geq0$, for all $v\in \mathbb{N}$, thus concluding the proof.
		\end{IEEEproof}
	
	\begin{Remark}[Novelty]\label{generaN}\emph{{
Our distributed dynamic periodic ETS \eqref{sys:trigger} was inspired by the dynamic ETSs in e.g., \cite{Girard2015,Deng2021,MISHRA2021ET} and the sampling-based ETSs in \cite{GUO2014}. Imitating the dynamic ETSs, an extra dynamic variable $\eta_i(\tau_v^i)>0$ was introduced in \eqref{sys:trigger} to relax
the triggering condition
from  $\rho_i(\tau_v^i)<0$ to $\rho_i(\tau_v^i)<-\frac{1}{\theta_i}\eta_i(\tau_{v}^{i})$
with $\theta_i>0$. Thus, a larger minimum transmission interval can be ensured by the ETS \eqref{sys:trigger} compared to the static ones in \cite{GUAN2020,SHI2019} that only exploit  the local sampled state $x_i(\tau_v^i)$ and the error $e_i(\tau_v^i)$.
Further, accounting for the relative state information $x_i(\tau_v^i)-x_j(\tau_v^i)$ employed by the distributed ETS \cite{GUO2014}, it relaxes the constraints $e_i^\top (\tau_v^i)\Omega_i e_i(\tau_v^i)>\sigma_1^i x_i^\top (\tau_v^i)\Omega_i x_i(\tau_v^i)$ on $e_i^\top (\tau_v^i)$ in decentralized ETSs \cite{GUAN2020,SHI2019} to $e_i^\top (\tau_v^i)\Omega_i e_i(\tau_v^i)>\sigma_1^i x_i^\top (\tau_v^i)\Omega_i x_i(\tau_v^i)
		+\sum_{j\neq i}^N \sigma_2^{ij} \left[x_i(\tau_v^i)-x_j(\tau_v^i)\right]^\top \Omega_i \left[x_i(\tau_v^i)-x_j(\tau_v^i)\right]$ in \eqref{sys:trigger}.
In Table \ref{Tab:compare} and Figs. \ref{FIG:x1}, \ref{FIG:x:data:dec}, and \ref{FIG:x:data:dis} of Section \ref{sec:dec:dis:compare}, a simulation example validates that the ETS \eqref{sys:trigger} generates less transmissions than the decentralized ETS (cf. \eqref{sys:trigger} with $\sigma_2^{ij}=0$) and the distributed ETS (cf. \eqref{sys:trigger} with $\sigma_1^{i}=0$), while maintaining similar performance.
On the other hand, as in the sampling-based ETSs, the proposed  transmission scheme \eqref{sys:trigger} only involves discrete-time signals at sampling instants $\tau_v^i$. When compared to discrete-time dynamic ETS \cite{MISHRA2021ET} that needs monitoring the state at each discrete point $t$, our scheme incurs a lower computational overhead if the sampling interval is determined as $h>1$.
Besides, different from the continuous-time dynamic periodic ETSs \cite{Girard2015,Deng2021}, the variable $\eta_i(\tau_v^i)$ evolves only successively at discrete sampling
points $\tau_v^i$ in the event-generator, which mitigates the
computational burden caused by continuously computing.
			In general, our ETS subsumes the discrete-time dynamic ETS \cite{MISHRA2021ET} (cf. \eqref{sys:trigger} with $h=1$ and $\sigma_1^{i}=0$), the static decentralized periodic ETS \cite{GUAN2020} (cf. \eqref{sys:trigger} with $\sigma_2^{ij}=0$ and $\theta$ goes to $\infty$), the static decentralized ETS \cite{SHI2019} (cf. \eqref{sys:trigger} with $h=1$, $\sigma_2^{ij}=0$, and $\theta\rightarrow \infty$), and the classic distributed ETS in \cite{GUO2014} (cf. \eqref{sys:trigger} with $\sigma_1^{i}=0$ and $\theta\rightarrow \infty$) as special cases. Thus, our triggering scheme
			in \eqref{sys:trigger} is expected to save transmission resources when compared to those ETSs.}}
	\end{Remark}
	
	\subsection{Model-based stability analysis}\label{sec:modle:stability}
	\begin{Theorem}[Model-based stability criterion]\label{Th1}
		For any scalars $\sigma_{1}^i\geq 0$, $\sigma_{2}^{ij}\geq 0$, $\bar{h}\geq\underline{h}\geq1$, and parameters $\lambda_i>0$, {$\theta_i> 0$} satisfying $1-\lambda_i-\frac{1}{\theta_i}\geq0$ for all $i\in\mathbb{N}_{[1,N]}$ and $j\in \mathcal{N}$, {asymptotic stability} of the  system \eqref{sys:sampling} is achieved under the triggered condition \eqref{sys:trigger}, and $\eta_i(\tau_v^{i})$ converges to the origin for any $\eta_i(0)\ge0$, if there exist matrices  $R_1\succ0$, $R_2\succ0$, $P\succ0$,
		$S$, $M_1$, $M_2$, $F$, and $\Omega_i\succ0$ for all $i\in\mathbb{N}_{[1,N]}$, satisfying the following LMIs $\forall h\in\{\underline{h},\bar{h}\}$
		\begin{align}{\label {Th1:LMI1}}
			&\left[
			\begin{array}{ccc}
				\Xi_0+h\Xi_a+\Psi+\mathcal{Q}  & hM_1\\
				\ast & -h\mathcal{R}_1
			\end{array}
			\right]\prec0\\
			{\label {Th1:LMI2}}
			&\left[
			\begin{array}{ccc}
				\Xi_0+h\Xi_b+\Psi+\mathcal{Q}  & hM_2\\
				\ast & -h\mathcal{R}_2
			\end{array}
			\right]\prec0
		\end{align}
		where
		\begin{align*}
			\Xi_0&:={\rm Sym}\big\{\Pi_1^\top  S \Pi_2 - \Pi_3^\top  S \Pi_4 + M_1\Pi_9 + M_2 \Pi_{10} \big\}\\
			&~~~~+ (H_2 - H_1)^\top  (R_2 - R_1) (H_2 - H_1)\\
&~~~~+H_2^\top  P H_2 -  H_1^\top  P H_1 \\
			\Xi_a&:={\rm Sym}\big\{\Pi_5^\top  S \Pi_6 \big\}+(H_2-H_1)^\top  R_2(H_2-H_1)\\
			\Xi_b&:={\rm Sym}\big\{\Pi_7^\top  S \Pi_8 \big\}+(H_2-H_1)^\top  R_1(H_2-H_1)\\
			\Psi&:={\rm Sym}\big\{F(AH_1+BKH_{7}-H_{2})\big\}\\
			\mathcal{Q}&:= H_3^\top  \mathrm{\Omega}_a H_3+
			H_{3}^\top \mathrm{\Omega}_b H_{3} - (H_3-H_{7})^\top  \Omega_a(H_3-H_{7})\\
			\mathcal{R}_1&:={\rm blkdiag}\big\{R_1~ 3R_1\big\},~
			\mathcal{R}_2:= {\rm blkdiag}\big\{R_2~ 3R_2\big\}\cr
			\Omega_a&:={\rm blkdiag}\big\{\sigma_1^1\Omega_1~\sigma_1^2\Omega_2~\ldots~\sigma_1^N\Omega_N\big\}\\
			\Omega_b&:=\\
			&         \left[\!\!\!\begin{array}{cccc}
				\sum_{j\neq 1}^N \sigma_2^{1j}\Omega_1+\sigma_2^{j1}\Omega_j&\cdots& -\sigma_2^{1N}\Omega_1-\sigma_2^{N1}\Omega_N\\
				\ast&\ddots&\vdots\\
				\ast&\ast&\sum_{j=1}^{N-1} \sigma_2^{lj}\Omega_1+\sigma_2^{jN}\Omega_j\\
			\end{array}\!\!\!\right]\\
			\Pi_1&:=\left[H_3^\top ,\, H_{4}^\top ,\,H_{2}^\top -H_3^\top ,\,H_{5}^\top +H_2^\top -H_3^\top \right]^\top \\
			\Pi_2&:=\left[-H_3^\top ,\, -H_4^\top ,\, H_4^\top -H_2^\top , \,H_6^\top -H_1^\top -H_4^\top \right]^\top \\
			\Pi_3&:=\left[H_0^\top ,\, H_0^\top ,\, H_1^\top -H_3^\top , \,H_5^\top -H_3^\top \right]^\top \\
			\Pi_4&:=\left[H_0^\top , \,H_0^\top ,\, H_4^\top -H_1^\top ,\, H_6^\top -H_4^\top \right]^\top \\
			\Pi_5&:=\left[H_3^\top , \,H_4^\top ,\, H_0^\top ,\, H_5^\top \right]\\
			\Pi_6&:=\left[-H_3^\top , \,-H_4^\top ,\, H_1^\top -H_2^\top ,\, -H_1^\top \right]^\top \\
			\Pi_7&:=\left[H_3^\top , \,H_4^\top ,\, H_2^\top -H_1^\top ,\, H_2^\top \right]^\top \\
\Pi_8&:=\left[H_3^\top , \,H_4^\top ,\, H_0^\top ,\, H_6^\top \right]^\top\\
			\Pi_9&:=\left[H_1^\top -H_3^\top ,\, H_1^\top +H_3^\top -2H_5^\top \right]^\top \\
			\Pi_{10}&:=\left[H_4^\top -H_1^\top , \,H_4^\top +H_1^\top -2H_6^\top \right]^\top \\
			H_\iota&:=\left[0_{n\times (\iota-1)n}, \,I_n, \,0_{n\times (7-\iota)n} \right], \iota=1, 2,\ldots,7\\
			H_0&:=0_{n\times7n}.
		\end{align*}
	\end{Theorem}

	\begin{IEEEproof}
Consider the intervals $[\tau_v^i, \tau_{v+1}^i-1]$ for all $v\in\mathbb{N}_{[0,m_k^i]}$ that satisfy $[t_k^i,t_{k+1}^i-1]=\bigcup \limits_{v=0}^{v=m_k^i} [\tau_v^i, \tau_{v+1}^i-1]$. 
	We choose the following functional
	\begin{align}{\label {Th1:Vt}}
		V(t)=V_a(x(t))+V_d(x(t),t)
	\end{align}
	where $t\in \mathbb{N}_{[\tau_v^i, \tau_{v+1}^i-1]}$, $V_a(x(t))=x^\top (t)Px(t)$, and $P \succ 0$;
	moreover, a discrete-time analog of looped-functional $V_d(x(t),t)$ is designed as follows
	\begin{equation}\label{Th1:WN}
		\begin{aligned}
			&V_d(x(t),t)=2\phi_1^\top (t)S \phi_2(t)\\
			&+(\tau_{v+1}^i-t)\Bigg[\sum \limits_{s=\tau_v^i}^{t} y^\top (s)R_1y(s) -y^\top (t)R_1y(t) \Bigg]\\
			&+(t-\tau_v^i)\Bigg[\sum \limits_{s=t}^{\tau_{v+1}^i} y^\top (s)R_2y(s) -y^\top (t)R_2y(t) \Bigg]
		\end{aligned}
	\end{equation}
	where $S$, $R_1\succ 0$, $R_2\succ 0$, and,
 \begin{align*}
 y(s)&:=x(s+1)-x(s), ~\phi_0:=\Big[x^\top ({\tau_v^i}),\,x^\top ({\tau_{v+1}^i})\Big]^\top, \\
 \phi_1(t)&:=  \Big[ (t-\tau_v^i)\phi_0^\top , x^\top (t) \!- \!x^\top ({\tau_v^i}), \sum \limits_{s=\tau_v^i}^{t}x^\top (s) \!-\! x^\top (\tau_v^i) \Big]^\top,\\
\phi_2(t)&:=  \Big[(\tau_{v+1}^i-t)\phi_0^\top , x^\top (\tau_{v+1}^i)-x^\top (t),\ldots\\
&~~~~~~~~~~~~~~~~~~~~~~~~~~~~~\ldots,\sum \limits_{s=t}^{\tau_{v+1}^i}x^\top (s)-x^\top (\tau_{v+1}^i)\Big]^\top .
\end{align*}

	{
	The forward difference of $V(t)$ yields that
	\begin{align}{\label {Th1:Vd}}
	\Delta V(t)= \Delta V_a(t)+ \Delta V_d(t)
	\end{align}
	where
	\begin{align*}
	\Delta V_a(t)&=\xi^\top (t)\Big(H_2^\top  P H_2 - H_1^\top  P H_1 \Big)\xi(t),\\
	\Delta V_{d}(t)&=\xi^\top (t)\Big[ (H_2 - H_1)^\top  (R_2 - R_1) (H_2 - H_1)\\
&~~~~~~~~~~~+2\Pi_1^\top  S \Pi_2 - 2\Pi_3^\top  S \Pi_4\\
&~~~~~~~~~~~+(t-\tau_{v}^i)\Xi_a+(\tau_{v+1}^i-t)\Xi_b\Big]\xi(t)\\
	&~~~-\sum \limits_{s=\tau_v^i}^{t-1} y^\top (s)R_1y(s)-\sum \limits_{s=t}^{\tau_{v+1}^i-1} y^\top (s)R_2y(s),
	\end{align*}
where $\xi(t):=\Big[x^\top (t)$, $x^\top (t+1)$, $x^\top (\tau_v^i)$, $x^\top (\tau_{v+1}^i)$, $\sum \limits_{s=\tau_v^i}^{t}\frac{x^\top (i)}{t-\tau_v^i+1}$,  $\sum \limits_{s=t}^{\tau_{v+1}^i}\frac{x^\top (i)}{\tau_{v+1}^i-t+1}$, $x^\top (t_k)\Big]^\top $.

Using the summation inequality \cite[Lemma 2]{Chen2017summation} with $N=1$, the summation terms satisfy
	\begin{equation}\label{th1:inequality}
	\begin{aligned}
	-&\sum \limits_{s=\tau_v^i}^{t-1} y^\top (s)R_1y(s)-\sum \limits_{s=t}^{\tau_{v+1}^i-1} y^\top (s)R_2y(s)\leq\\
	&~~\xi^\top (t)\Big[(t-\tau_v^i)M_1 \mathcal{R}_1^{-1}M_1^{\top}+2M_1\Pi_9\\
	&~~~~~~~~~+(\tau_{v+1}^i-t)M_2 \mathcal{R}_2^{-1}M_2^{\top}+2M_2\Pi_{10}\Big]\xi(t)
	\end{aligned}
	\end{equation}
	Summing up \eqref{Th1:Vd}-\eqref{th1:inequality}, we arrive at}
	\begin{equation}{\label {Th1:sum}}
		\begin{aligned}
			\Delta V(t)\leq&~\xi^{\top}(t)\Big[\Xi_0+(t-\tau_{v}^i)\big(\Xi_a+M_1 \mathcal{R}_1^{-1} M_1^\top \big)\\
			&~~~~~~~+(\tau_{v+1}^i-t)\big(\Xi_b+M_2 \mathcal{R}_2^{-1} M_2^\top \big)\Big]\xi(t)
		\end{aligned}
	\end{equation}
	
	Through the descriptor method \cite{Fridman2010}, the model-based system representation \eqref{sys:sampling} can be written as follows
	\begin{align}{\label {Th1:zero}}
		0&=2\xi^\top (t)F \big[A x(t)+BK{x_k(t)}- x(t+1) \big] \notag\\
		&=2\xi^\top (t)F \big(AH_1+BKH_{7}-H_{2}\big)\xi(t),~ \forall t \in \mathbb{N}_{[\tau_v^i, \tau_{v+1}^i-1]}.
	\end{align}
	
	In light of \eqref{sys:trigger}, Lemma \ref{lemma:nonneg.dynam} asserts that $\eta_i(\tau_v^i)\geq0$ for $\lambda_i>0$, $\eta_i(0)\ge0$, and $1-\lambda_i-\frac{1}{\theta_i}\geq0$ when
	$\theta\neq 0$. Hence, it holds that $\sum_{i=1}^N \eta_i(\tau_v^i)\geq0$.
	Calculating the forward difference of $\sum_{i=1}^N \eta_i(\tau_v^i)$ and according to \eqref{sys:dynamic}, it yields that
		\begin{align}
			\sum_{i=1}^N [\eta_i(\tau_{v+1}^i)-\eta_i(\tau_{v}^i)]&=\sum_{i=1}^N[-\lambda_i\eta_i(\tau_v^{i})+\rho_i(\tau_v^{i})]\nonumber\\
			&=\xi^\top (t)\mathcal{Q}\xi(t)-\sum_{i=1}^N\lambda_i\eta_i(\tau_v^{i})\nonumber\\
&			\leq \xi^\top (t)\mathcal{Q}\xi(t).\label{th1:eta:sum:ineuqlity}
		\end{align}

	Combining \eqref{Th1:sum} with \eqref{th1:eta:sum:ineuqlity}, we have that
		\begin{align}\label{Th1:vt:dt}
			\Delta V(t)&+\sum_{i=1}^N[\eta_i(\tau_{v+1}^i)-\eta_i(\tau_{v}^i)]\leq\nonumber\\ &\xi^{\top}(t)\left[\frac{t-\tau_v^i}{h}\Upsilon_1(h) +\frac{\tau_{v+1}^i-t}{h}\Upsilon_2(h) \right]\xi(t)
		\end{align}
	where 
	\begin{equation*}
	\begin{aligned} \Upsilon_1(h)&=\Xi_0+\Psi+\mathcal{Q}+h\Xi_{a} +h M_1 \mathcal{R}_1^{-1} M_1^\top,\\	\Upsilon_2(h)&=\Xi_0+\Psi+\mathcal{Q}+h\Xi_{b} +h M_2 \mathcal{R}_2^{-1} M_2^\top .
	\end{aligned}
	\end{equation*}
	
	According to the Schur Complement Lemma, inequalities $\Upsilon_1(h)\prec0$ and $\Upsilon_2(h)\prec0$ are equivalent
	to LMIs \eqref{Th1:LMI1} and \eqref{Th1:LMI2}, which are affine, and  convex, with respect to $h$. Thus, \eqref{Th1:LMI1} and \eqref{Th1:LMI2} at the vertices of $h\in[\underline{h},\bar{h}]$ ensure $\Delta V(t)+\sum_{i=1}^N [\eta_i(\tau_{v+1}^i)-\eta_i(\tau_{v}^i)]<0$ for all $h\in[\underline{h},\,\bar{h}]$.
	It follows from the looped-functional  approach \cite{SEURET2012177} that
	\begin{equation}{\label {Th1:vj}}
		\begin{aligned}
			 \sum_{s=\tau_{v}^i}^{\tau_{v+1}^i-1}\Big\{\Delta V(s)+\sum_{i=1}^N [\eta_i(\tau_{v+1}^i)-\eta_i(\tau_{v}^i)]\Big\}< 0
		\end{aligned}
	\end{equation}
	which ensures that $V_a(\tau_{v+1}^i)+(h-1)\sum_{i=1}^N \eta_i(\tau_{v+1}^i)
	<
	V_a(\tau_{v}^i)
	+(h-1)\sum_{i=1}^N  \eta_i(\tau_{v}^i), ~\forall x(\tau_{v}^i)\neq 0$.
	We conclude that the state of system \eqref{sys:sampling} and $\eta_i(\tau_v^{i})$ converge to the origin under our
	transmission scheme, which completes the proof.
		\end{IEEEproof}
	

{
\begin{Remark}[Discrete-time looped-functional]
\emph{Motivated by the continuous-time looped-functional approach in \cite{SEURET2012177,Wang2021} that has been
 used for single sampled-data control systems, a discrete-time looped-functional
 is constructed in \eqref{Th1:WN} for network systems containing multiple subsystems. It can be easily proved that the proposed looped-functional $V_d(x(t),t)$ has the feature of
 $V_d(x(\tau_v^i),\tau_v^i)=V_d(x(\tau_{v+1}^i),\tau_{v+1}^i)$, which essentially relaxes the requirement of  the common Lypapunov functional $V_a(x(t))$ decreasing at each discrete time $t$, but necessarily ensures that $V_a(x(t))$ descends at designed sampling points $\tau_v^i$, cf. in \eqref{Th1:vj}. Thus, less conservative stability conditions can be obtained by using the looped-functional approach when compared to the traditional discrete-time Lyapunov functional. Besides,
obtaining a sampling-dependent condition
is another reason why we introduce the looped-functional. An allowable
sampling interval can be searched for by using LMIs \eqref{Th1:LMI1}-\eqref{Th1:LMI2}.}
\end{Remark}}

{\subsection{Model-based controller design}\label{design:event:model}
Theorem \ref{Th1} only provides an analysis method for event-triggered systems.
We cannot  obtain the controller gain $K$ directly from solving the LMIs \eqref{Th1:LMI1} and \eqref{Th1:LMI2}, due to that $K$ is coupled with the free matrix $F$.
This part gives a new model-based condition  for co-designing the distributed controllers $K_i$, $K_{ij}$, and the ETS parameters based on Theorem \ref{Th1} while guaranteeing the stability. To this end, we start with an equivalently transformed system from \eqref{sys:sampling}.
	Assume that $G_i\in \mathbb{R}^{n \times n}$ are nonsingular, and let $x_i(t)=G_iz_i(t)$. The system \eqref{sys:sampling} is restructured as follows
	\begin{equation}\label{Design:NCS}
		z(t+1)=G^{-1}AG z(t)+G^{-1}B K_c z(t_k),~t\in \mathbb{N}_{[t_k^i, t_{k+1}^i-1]}
	\end{equation}
	where $z(t):=[z_1^\top (t)~z_2^\top (t)\cdots~z_N^\top (t)]^\top $, $K_c:=KG$, and $G:={\rm blkdiag}\{G_1~G_2~\cdots~G_N\}$.
	The system \eqref{Design:NCS} performs the same stability behavior as system \eqref{sys:sampling}, since they are equivalent algebraically. Imitating Theorem \ref{Th1}, the following stability criterion can be derived using the equivalent system \eqref{Design:NCS}.

\begin{Theorem}
		[Model-based co-designing controller and triggering matrix]\label{Th:design:model}
		For any scalars $\sigma_{1}^i\geq 0$, $\sigma_{2}^{ij}\geq 0$, $\bar{h}\geq\underline{h}\geq1$, and parameters $\lambda_i>0$, $\theta_i> 0$ obeying $1-\lambda_i-\frac{1}{\theta_i}\geq0$ for all $i\in\mathbb{N}_{[1,N]}$ and $j\in \mathcal{N}$, there exists a block controller gain $K$ such that asymptotic stability of the system \eqref{sys:sampling} is achieved
under the transmission scheme \eqref{sys:trigger}, and $\eta_i(\tau_{v}^i)$ converges to the origin for any $\eta_i(0)\ge0$,
if there exist matrices $R_1\succ0$, $R_2\succ0$, $P\succ0$,  $\Omega^z_{i}\succ0$,
		$S$, $M_1$, $M_2$, $G$, $K_c$, and a scalar $\varepsilon>0$, satisfying the following LMIs $\forall h\in\{\underline{h}, \bar{h}\}$
\begin{align}{\label {Th2:LMI1}}
			&\left[
			\begin{array}{ccc}
				\Xi_0+h\Xi_a+\hat{\Psi}+\mathcal{Q}^z  & hM_1\\
				\ast & -h\mathcal{R}_1
			\end{array}
			\right]\prec0\\
			{\label {Th2:LMI2}}
			&\left[
			\begin{array}{ccc}
				\Xi_0+h\Xi_b+\hat{\Psi}+\mathcal{Q}^z  & hM_2\\
				\ast & -h\mathcal{R}_2
			\end{array}
			\right]\prec0
		\end{align}
where 
\begin{align*}
\mathcal{D}&:=(H_1+2 H_2)^\top\\
\hat{\Psi}&:={\rm Sym}\big\{\mathcal{D}(AGH_1+BK_cH_{7}-GH_{2})\big\}\\
\mathcal{Q}^z&:= H_3^\top  \mathrm{\Omega}^z_a H_3+
			H_{3}^\top \mathrm{\Omega}^z_b H_{3} - (H_3-H_{7})^\top  \Omega^z_a(H_3-H_{7})\\
			\Omega_a^z&:={\rm blkdiag}\big\{\sigma_1^1\Omega_1^z~\sigma_1^2\Omega_2^z~\ldots~\sigma_1^N\Omega_N^z\big\}\\
			\Omega_b^z&:=\\
			&         \left[\!\!\!\begin{array}{cccc}
				\sum_{j\neq 1}^N \sigma_2^{1j}\Omega_1^z+\sigma_2^{j1}\Omega_j^z&\cdots& -\sigma_2^{1N}\Omega_1^z-\sigma_2^{N1}\Omega_N^z\\
				\ast&\ddots&\vdots\\
				\ast&\ast&\sum_{j=1}^{N-1} \sigma_2^{lj}\Omega_1^z+\sigma_2^{jN}\Omega_j^z\\
			\end{array}\!\!\!\right]\\
\end{align*}
Moreover, the desired block controller gain and the triggering matrices are given by $K=K_cG^{-1}$, $\Omega_a={G^{-1}}^\top{\Omega}_a^z{G^{-1}}$, and $\Omega_b={G^{-1}}^\top{\Omega}_b^z{G^{-1}}$, respectively.
\end{Theorem}

\begin{IEEEproof}
We construct the following functional $V_z(t)$ by replacing $x(t)$ of the functional $V(t)$ in \eqref{Th1:Vt} with $z(t)$ in \eqref{Design:NCS}
\begin{align}{\label {Th2:Vt}}
		V_z(t)=V_a(z(t))+V_d(z(t),t)
	\end{align}
	where $V_a(z(t))=z^\top (t)Pz(t)$ and
	the looped-functional $V_d(z(t),t)$ is
	formulated according to \eqref{Th1:WN}.

Besides, it follows from the system equation
 in \eqref{Design:NCS} that
 \begin{align}{\label {Th2:zero}}
		0&=2[z(t)+2 z(t+1)]^\top G \\
&~~~\times \big[G^{-1}AG z(t)+G^{-1}B K_c z(t_k)-z(t+1)\big] \notag\\
		&=2\xi_z^\top (t) \big[\mathcal{D}(AGH_1+BK_cH_{7}-GH_{2})\big]\xi_z(t)
	\end{align}
where $\xi_z(t):=[z^\top (t)$, $z^\top (t+1)$, $z^\top (\tau_v^i)$, $z^\top (\tau_{v+1}^i)$, $\sum \limits_{s=\tau_v^i}^{t}\frac{z^\top (i)}{t-\tau_v^i+1}$,  $\sum \limits_{s=t}^{\tau_{v+1}^i}\frac{z^\top (i)}{\tau_{v+1}^i-t+1}$, $z^\top (t_k)]^\top $.

Recalling \eqref{th1:eta:sum:ineuqlity}, the triggering condition \eqref{sys:trigger} ensures
$\sum_{i=1}^N [\eta_i(\tau_{v+1}^i)-\eta_i(\tau_{v}^i)]\leq \xi^\top (t)\mathcal{Q}\xi(t)$. Then,
the following inequality holds with $x_i(t)=G_iz_i(t)$ and $\xi^{\top}(t)\mathcal{Q}\xi(t)=\xi_z^{\top}(t)\mathcal{Q}^z\xi_z(t)$
\begin{equation}
\sum_{i=1}^N [\eta_i(\tau_{v+1}^i)-\eta_i(\tau_{v}^i)]\leq \xi_z^\top (t)\mathcal{Q}^z\xi_z(t).
\end{equation}
Similar to \eqref{Th1:vt:dt}, it can be derived that
\begin{align}\label{Th2:vt:dt}
			\Delta V_z(t)&+\sum_{i=1}^N[\eta_i(\tau_{v+1}^i)-\eta_i(\tau_{v}^i)]\leq\nonumber\\ &\xi_z^{\top}(t)\left[\frac{t-\tau_v^i}{h}\hat{\Upsilon}_1(h) +\frac{\tau_{v+1}^i-t}{h}\hat{\Upsilon}_2(h) \right]\xi_z(t)
		\end{align}
where
\begin{equation*}
	\begin{aligned} \hat{\Upsilon}_1(h)&=\Xi_0+\hat{\Psi}+\mathcal{Q}^z+h\Xi_{a} +h M_1 \mathcal{R}_1^{-1} M_1^\top,\\	\hat{\Upsilon}_2(h)&=\Xi_0+\hat{\Psi}+\mathcal{Q}^z+h\Xi_{b} +h M_2 \mathcal{R}_2^{-1} M_2^\top .
	\end{aligned}
	\end{equation*}
Finally, similar to the proof of Theorem \ref{Th1}, the LMIs \eqref{Th2:LMI1} and \eqref{Th2:LMI2} can be shown equivalent to $\hat{\Upsilon}_1(h)\prec0$ and $\hat{\Upsilon}_2(h)\prec0$, respectively, which ensure asymptotic stability of the systems \eqref{Design:NCS} and $\eta_i(\tau_{v}^i)$. System \eqref{Design:NCS} exhibits the same dynamic behavior and stability properties as \eqref{sys:sampling}, as $G$ is nonsingular, which ends the proof.
\end{IEEEproof}
}	
	\subsection{Data-driven controller design}\label{design:event}
	We are now ready to provide a data-driven method for co-designing a full structure of the block controller gain $K$, and all matrices $\Omega_i$ of the distributed triggering scheme  for system \eqref{sys:sampling} with \emph{unknown} matrices $A_i$, $B_i$, $A_{ij}$.
	Motivated by \cite{Berberich2020,wildhagen2021datadriven},
	the core idea is to replace the model in \eqref{sec1:sys:disLTI} with a data-driven system expression only using the measurements $\{x_i(T)\}^{\rho}_{T=0}$ and $\{u_i(T)\}^{\rho-1}_{T=0}$. 
	Following this line, a data-based stability condition is obtained by combining
	the data-driven parametrization in Lemma \ref{Lemma:system:data} and the stability guarantee in
 Theorem \ref{Th:design:model}.
	\begin{Theorem}
		[Data-driven co-designing controller
				and triggering matrix]\label{Th:data}
		For any scalars $\sigma_{1}^i\geq 0$, $\sigma_{2}^{ij}\geq 0$, $\bar{h}\geq\underline{h}\geq1$, and parameters $\lambda_i>0$, {$\theta_i> 0$} obeying $1-\lambda_i-\frac{1}{\theta_i}\geq0$ for all $i\in\mathbb{N}_{[1,N]}$ and $j\in \mathcal{N}$, there exists a block controller gain $K$ such that {asymptotic stability} of system \eqref{sys:sampling} is achieved
under the transmission scheme \eqref{sys:trigger} for any $[A ~B]\in \Sigma_{AB}$, and $\eta_i(\tau_{v}^i)$ converges to the origin for any $\eta_i(0)\ge0$,
if there exist matrices $R_1\succ0$, $R_2\succ0$, $P\succ0$,  $\Omega^z_{i}\succ0$,
		$S$, $M_1$, $M_2$, $G$, $K_c$, and a scalar $\varepsilon>0$, satisfying the following LMIs $\forall h\in\{\underline{h}, \bar{h}\}$
		\begin{align}{\label {Th:data:LMI1}}
			&\left[
			\begin{array}{ccc}
				\mathcal{T}_1& \mathcal{F}+\mathcal{T}_2& 0\\
				\ast &  \Xi_0+h\Xi_a+\bar{\Psi}+\mathcal{Q}^z+\mathcal{T}_3  & hM_1\\
				\ast &  \ast & -h\mathcal{R}_1
			\end{array}
			\right]\prec0\\
			{\label {Th:data:LMI2}}
			&\left[
			\begin{array}{ccc}
				\mathcal{T}_1& \mathcal{F}+\mathcal{T}_2& 0\\
				\ast &  \Xi_0+h\Xi_b+\bar{\Psi}+\mathcal{Q}^z+\mathcal{T}_3  & hM_2\\
				\ast & \ast & -h\mathcal{R}_2
			\end{array}
			\right]\prec0
		\end{align}
where
		\begin{align*}
\bar{\Psi}&:={\rm Sym}\big\{\!-\mathcal{D}GH_{2}\big\},
~\mathcal{F}:=\!\Big[H_1^\top G^\top,~ H_{7}^\top K_c^\top \Big]^\top \\
			\mathcal{V}_1&:=
			\left[\begin{array}{ccc}I & 0\\\end{array}\right],~
			\mathcal{V}_2:=
			\left[\begin{array}{ccc} 0& \mathcal{D}\\\end{array}\right]\\
\mathcal{T}_1&:=\varepsilon\mathcal{V}_1\Theta_{AB}\mathcal{V}_1^\top ,~
			 \mathcal{T}_2:=\varepsilon\mathcal{V}_1\Theta_{AB}\mathcal{V}_2^\top ,~
			 \mathcal{T}_3:=\varepsilon\mathcal{V}_2\Theta_{AB}\mathcal{V}_2^\top .
		\end{align*}
Moreover, the desired block controller gain and the triggering matrices are given by $K=K_cG^{-1}$, $\Omega_a={G^{-1}}^\top{\Omega}_a^z{G^{-1}}$, and $\Omega_b={G^{-1}}^\top{\Omega}_b^z{G^{-1}}$, respectively.
	\end{Theorem}
	
		\begin{IEEEproof}
{
	According to \eqref{Th2:vt:dt} in the proof of Theorem \ref{Th:design:model}, 
$\hat{\Upsilon}_1(h)\prec0$ and $\hat{\Upsilon}_2(h)\prec0$ are sufficient conditions guaranteeing
the asymptotic stability of system \eqref{sys:sampling}.
	For deriving model-based condition, we restructure $\hat{\Upsilon}_1(h)$ and $\hat{\Upsilon}_2(h)$ as follows}
	\begin{align*}
		\hat{\Upsilon}_1(h)&:=
		 \left[\begin{array}{cc}[\mathcal{D}A~\mathcal{D}B]^{\top}\\I \\\end{array}\right]^{\top}
		\left[\begin{array}{cc}0 & \mathcal{F}\\\ast & \hat{\Upsilon}_1(h)-\hat{\Psi} +\bar{\Psi} \\\end{array}\right]
		\left[\cdot\right]\\
\hat{\Upsilon}_2(h)&:=
		 \left[\begin{array}{cc}[\mathcal{D}A~\mathcal{D}B]^{\top}\\I \\\end{array}\right]^{\top}
		\left[\begin{array}{cc}0 & \mathcal{F}\\\ast & \hat{\Upsilon}_2(h)-\hat{\Psi} +\bar{\Psi} \\\end{array}\right]
		\left[\cdot\right].
	\end{align*}

{In the light of the data-based representation in Lemma \ref{Lemma:system:data}, it holds for any $[A ~ B]\in\Sigma_{AB}$ that
	\begin{equation}
	\begin{aligned}
	\left[ \begin{array}{cc}[A~B]^{\top}\\I \\\end{array} \right]^{\top}
	   \Theta_{AB}
	  \left[ \begin{array}{cc}[A~B]^{\top} \\
	   I\\\end{array} \right]\succeq0.
	\end{aligned}
	\end{equation}}
	
	Using the full-block S-procedure \cite{Sche2001}, inequalities $\hat{\Upsilon}_1(h)\prec0$ and $\hat{\Upsilon}_2(h)\prec0$ are ensured for any $[A ~B]\in\Sigma_{AB}$ if there is some $\varepsilon>0$ adhering to the following LMIs
	\begin{align}\label{Th:data:fullblock1}
		&\left[\!\!\begin{array}{cc}0 & \mathcal{F}\\\ast & \hat{\Upsilon}_2(h)-\hat{\Psi} +\bar{\Psi} \\\end{array}\!\!\right]+
		\varepsilon \left[\!\!\begin{array}{cc}\mathcal{V}_1\Theta_{AB}\mathcal{V}_1^{\top} & \mathcal{V}_1\Theta_{AB}\mathcal{V}_2^{\top}\\\ast &  \mathcal{V}_2\Theta_{AB}\mathcal{V}_2^{\top} \\\end{array}\!\!\right]
		\prec0\\
\label{Th:data:fullblock2}
&\left[\!\!\begin{array}{cc}0 & \mathcal{F}\\\ast & \hat{\Upsilon}_2(h)-\hat{\Psi} +\bar{\Psi} \\\end{array}\!\!\right]+
		\varepsilon \left[\!\!\begin{array}{cc}\mathcal{V}_1\Theta_{AB}\mathcal{V}_1^{\top} & \mathcal{V}_1\Theta_{AB}\mathcal{V}_2^{\top}\\\ast &  \mathcal{V}_2\Theta_{AB}\mathcal{V}_2^{\top} \\\end{array}\!\!\right]
		\prec0
	\end{align}

	The Schur Complement Lemma guarantees that the inequalities in \eqref{Th:data:fullblock1} and \eqref{Th:data:fullblock2} are equivalent to \eqref{Th:data:LMI1} and \eqref{Th:data:LMI2}. Subsequently, we can draw a conclusion that \eqref{Th:data:LMI1} and \eqref{Th:data:LMI2} at the vertices of $h\in[\underline{h},\bar{h}]$ ensure asymptotic stability of system \eqref{sys:sampling}
	under the condition \eqref{sys:trigger} for any
	$[A ~B]\in \Sigma_{AB}$, and $\eta_i(\tau_{v}^i)$ in \eqref{sys:trigger} also converges to the origin.\end{IEEEproof}
	{
	\begin{Remark}[Conservatism]
	\emph{
Theorem \ref{Th:data} offers a data-based stability criterion for unknown sampled-data control system \eqref{sys:sampling}, by which a allowable  sampling interval $h$ can be
searched for only using state-input data.
	In our sampling-based ETS \eqref{sys:trigger}, a larger $h$ may lead to a lower communication rate thus to save network resources. For this purpose,
	three methods can be taken into consideration for reducing the conservatism of the data-based condition (cf. Theorem \ref{Th:data}), which is beneficial for obtaining a greater $h$.
Firstly, selecting a suitable looped-functional that considers more system information is helpful for decreasing the conservatism of Theorem \ref{Th:data}, due to the fact that Theorem \ref{Th:data} is built based on the model-based condition in Theorem \ref{Th1}. Note that the proposed looped-functional  \eqref{Th1:WN} only employs the system states and their single summations. According to the latest research results in \cite{Wang2021,Park2020} for single systems, less conservative stability conditions may be obtained if higher-order summations of the system states are used in constructing a proper looped-functional.
Secondly, a tighter bounding technique used for estimating the nonlinear terms in \eqref{Th1:Vd} can enhance the feasibility of the condition; e.g.,  choosing the summation inequality \cite[Lemma 2]{Chen2017summation} in the sense of  $N>1$ instead of $N=1$ in this paper.  Thirdly, leveraging prior knowledge on the system matrices or on the disturbance in
data-driven system construction (cf. in Lemma \ref{Lemma:system:data}) can significantly improve the performance of our data-driven method. For example, in the same spirit of \cite{berberich2020combining} for single systems, a less conservative noise bound (cf. Assumption \ref{Ass:disturbance}) can be reformulated if we assume that the disturbance satisfies a pointwise-in-time Euclidean norm bound of the
form $\|w_i(T)\|\leq \bar{w}$ with $\bar{w}>0$ for all $T = 0, \ldots, \rho-1$.
The detailed research about this topic can be  found in \cite{berberich2020combining}.} 
	\end{Remark}

\begin{Remark}[Comparison]\label{Remark:conservatism}
\emph{Theorem \ref{Th:data} guarantees the asymptotic stability of all systems having $[A ~B]\in \Sigma_{AB}$. Compared
with the model-based condition in Theorem \ref{Th:design:model} for the unique $[A ~B]$, our data-driven condition is more conservative, thus to inevitably supply smaller room for optimizing the transmission frequency and the system performance.  
Such a conclusion is certificated in Section \ref{sec:example} through a numerical comparison between Figs. \ref{FIG:x1} and \ref{FIG:x1:model}, showing that our model-based
method (cf. Theorem \ref{Th:design:model}) ensures a lower transmission frequency than the data-based one (cf. Theorem \ref{Th:data}) with the same triggering parameters, but results in faster convergence rates of the system trajectories.
It  should also be mentioned that Theorem \ref{Th:data} offers a condition for the unknown unperturbed sampled-data system \eqref{sys:sampling}. Our future works will focus on extending it to the case of guaranteeing an $\mathcal{L}_2$-gain performance on the perturbed system with bounded noise.}
\end{Remark}}
	
{\begin{Remark}[Length of data]
\emph{With the observation of LMIs \eqref{Th:data:LMI1} and \eqref{Th:data:LMI2}, a  merit of  the data-driven method in Theorem \ref{Th:data} is that the dimensions of the corresponding LMIs are independent of the number of the collected state-input data (cf. $\{x_i(T)\}^{\rho}_{T=0}$ and $\{u_i(T)\}^{\rho-1}_{T=0}$). Different from
the data-driven control of \cite{persis2020},  our method is scalable to any length of data. Besides, it has been demonstrated in
\cite{berberich2020combining} that the guaranteed system performance may be improved
by increasing the data length $\rho$ in the data-driven control design, as more data
can provide more accurate system description in Lemma \ref{Lemma:system:data}. However, a higher computational complexity  arises  with the size of the data, which becomes challenging for large-scale network systems. This motivates us to balance the desired system performance and the amount of  data, which also constitutes our future research direction.}
\end{Remark}}

{\begin{Remark}
\emph{The distributed control strategy \eqref{feedbackC} and ETS \eqref{sys:trigger} only require local information of the system, i.e., the agent's and its neighbors' sampled states. However, data-driven control protocols in Theorems \ref{Th:data} are not fully distributed, which rely on the global information of the network graph, e.g., when constructing the data-driven system representation in Lemma \ref{Lemma:system:data}.
For  real network systems, it is possible to obtain the structure of the systems in advance and  collect all state-input data of the subsystems in offline. A practical
numerical example in Fig. \ref{FIG:inverted} is given to demonstrate the effectiveness of the data-driven control method.
}
\end{Remark}}

	\section{Examples and Simulation}\label{sec:example}
	
\begin{figure}[t]
		\centering
		\includegraphics[scale=0.8]{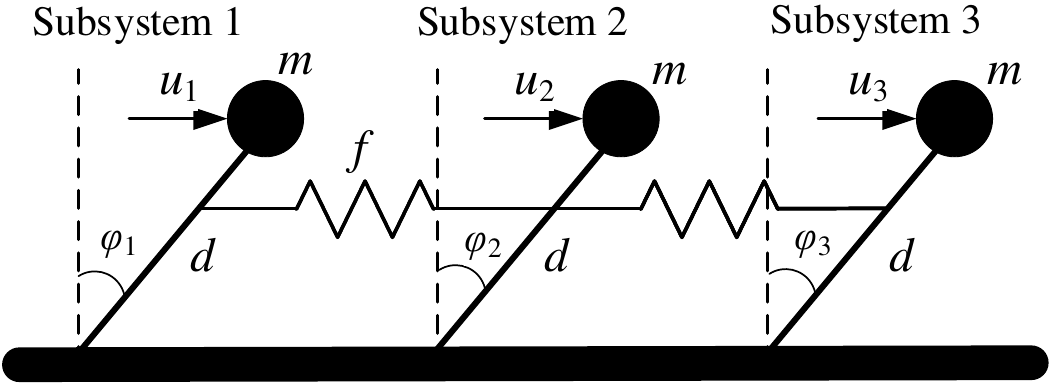}
		\caption{A network of coupled inverted pendulums.}
		\label{FIG:inverted}
	\end{figure}

A network system consists of three identical and coupled inverted pendulums  is employed in this section to examine  the proposed data-driven distributed event-triggered control methods; see Fig. \ref{FIG:inverted} for a pictorial description.
{Each  subsystem (inverted pendulum) has the following dynamics, for $i, j=1,2,3$ and $j\neq i$
\begin{equation*}
		\dot x_i(l)=\bar{A}_i x_i(l)+\sum_{j \in \mathcal{N}} \bar{A}_{ij}x_j(l)+\bar{B}_i u_i(l), ~l\geq 0
	\end{equation*}
where the system and coupling matrices are given as
			\begin{equation*}
				\bar{A}_i=\left[\!
				\begin{array}{cc}
					0& 1 \\
					\frac{g}{d}-\frac{N_if}{md^2} & 0 \\
				\end{array}
				\right],
				\bar{B}_i=\left[\!
				\begin{array}{cc}
					0 \\
					\frac{1}{md^2}\\
				\end{array}
				\right],
				\bar{A}_{ij}=\left[\!
				\begin{array}{cc}
					0& 0\\
					\frac{c_{ij}f}{md^2} & 0 \\
				\end{array}
				\right],
			\end{equation*}}
			and $g=10 m/s^2$ is the acceleration due to gravity, $m=1 kg$ is the mass of the pendulum, $d=2m$  is the length of the pendulum, and $f=5 N/m$ is the spring constant; and $N_i$ denotes
			the number of neighbors of the $i$-th subsystem.
The adjacency matrix $\mathcal{C}$ describing the communication graph
is given as follows
\begin{equation*}
\mathcal{C}=\left[
				\begin{array}{ccc}
					0& 1 & 0\\
					1& 0 &1\\
                    0& 1& 0
				\end{array}
				\right].
\end{equation*}

			
			Upon discretization, we arrive at the following discrete-time linear  system
			 \begin{equation}\label{example:discrete:system}
				x(t+1)=A(T_k) x(t)+B(T_k) u(t), ~t\in \mathbb{N}
			\end{equation}
			where $x(t):=[x_1^\top (t)~ x_2^\top(t)~\cdots~x_N^\top (t)]^\top $, $x_i(t)=[\varphi_i(t)~\, \dot\varphi_i(t)]^{\top}$, $T_k>0$ is the discretization interval,
			$A(T_k):=e^{AT_k}$, and
			$B(T_k):=\int_0^{T_k}e^{A(s)}{B}ds$.
			We consider using a distributed linear state-feedback controller \eqref{feedbackC} to control the system. System \eqref{example:discrete:system} is rewritten as
			 \begin{equation}\label{example:sample:system}
				x(t+1)=A(T_k) x(t)+B(T_k) K{x_k(t)}, ~t\in \mathbb{N}_{[t_k^i, t_{k+1}^i-1]}.
			\end{equation}
			
			The proposed data-driven ETS \eqref{sys:trigger} is then applied to system \eqref{example:sample:system}. In the following part, {we 
examine the effectiveness of the proposed data- and model-based co-design methods. All numerical computations were performed using Matlab, together with the SeDuMi toolbox \cite{SeDuMi}.
}

\begin{figure}[t]
		\centering
\subfigure{
\includegraphics[scale=0.55]{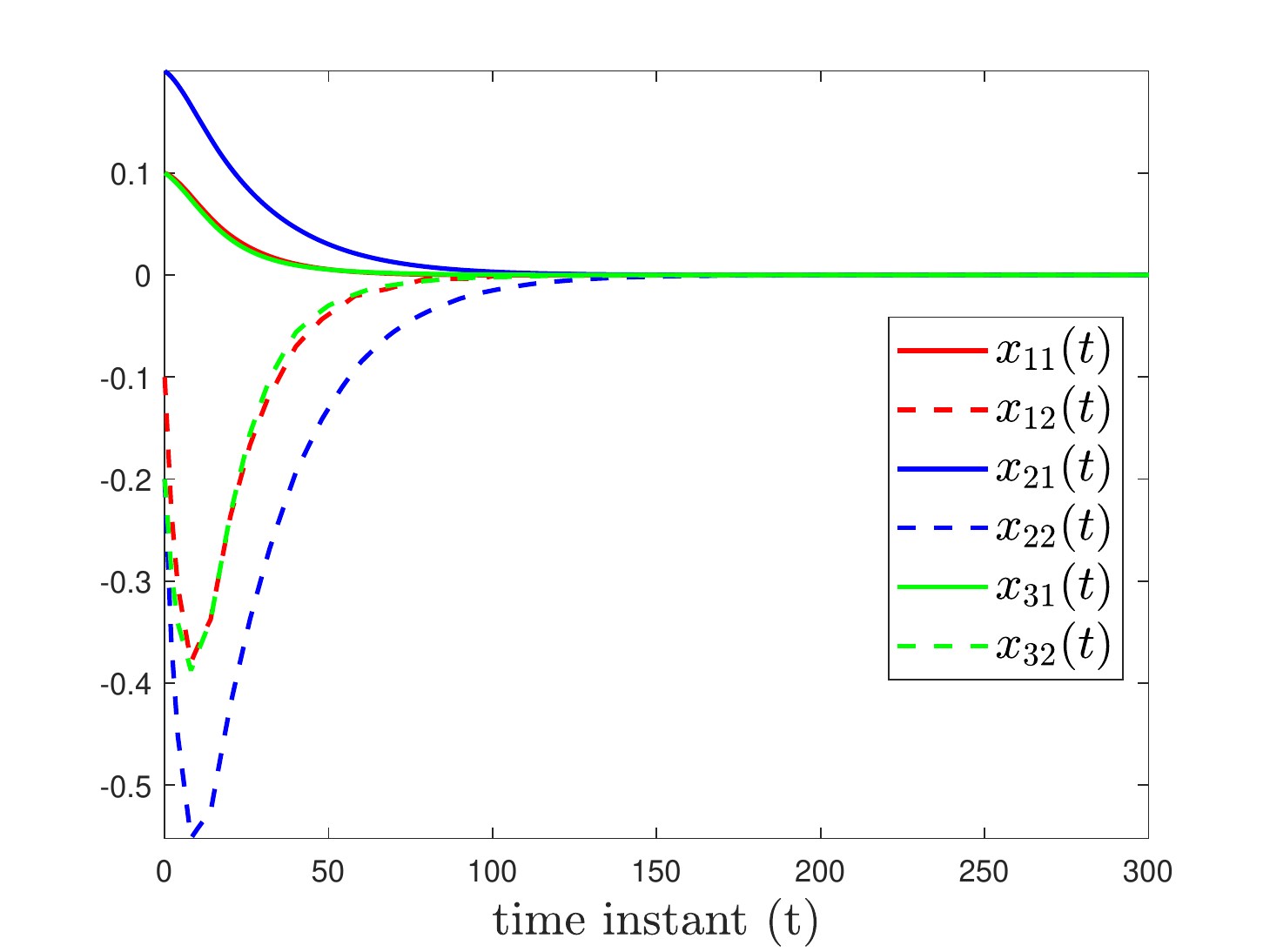}
}
\subfigure{
\includegraphics[scale=0.55]{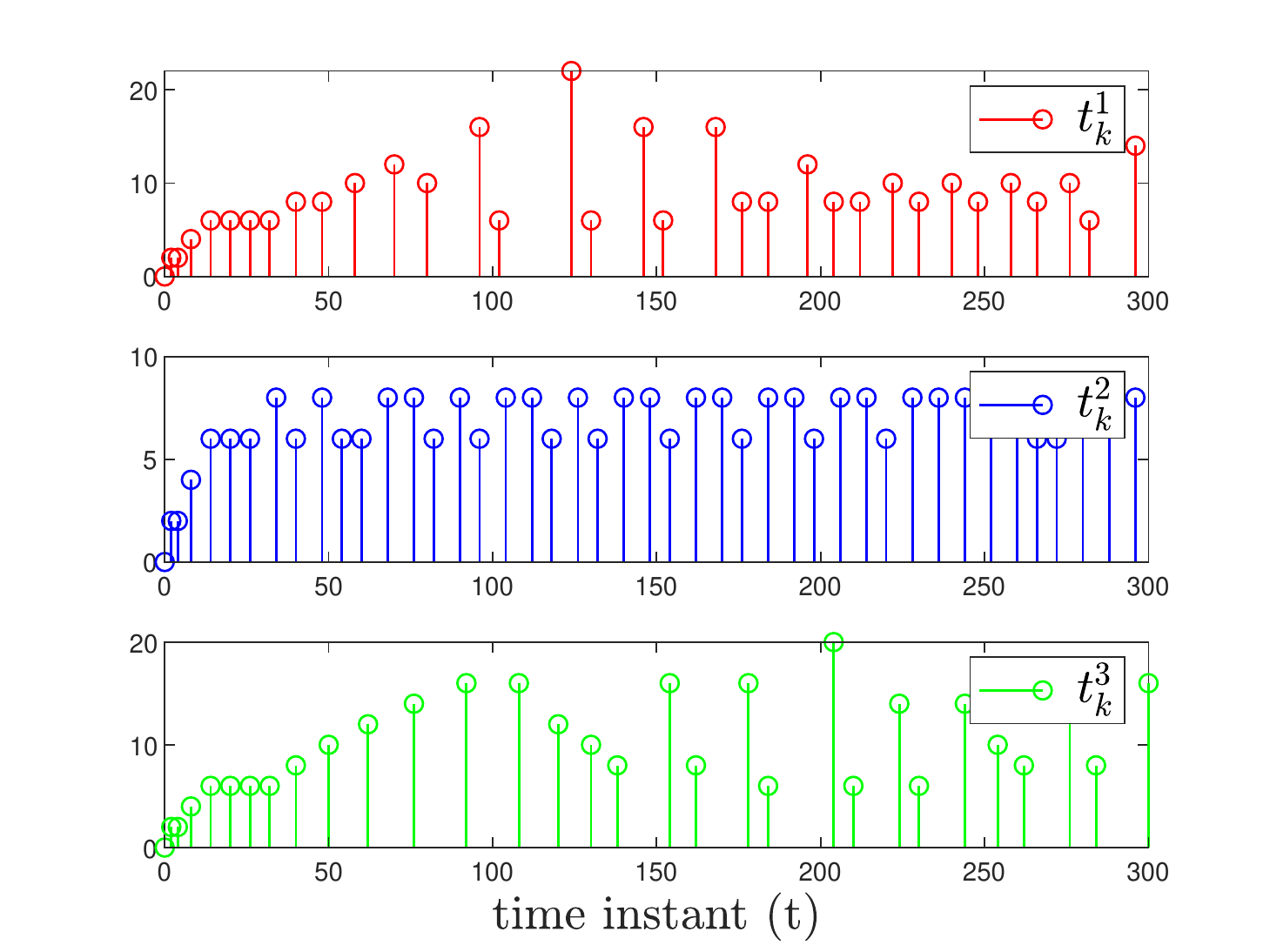}
}	
\subfigure{
\includegraphics[scale=0.55]{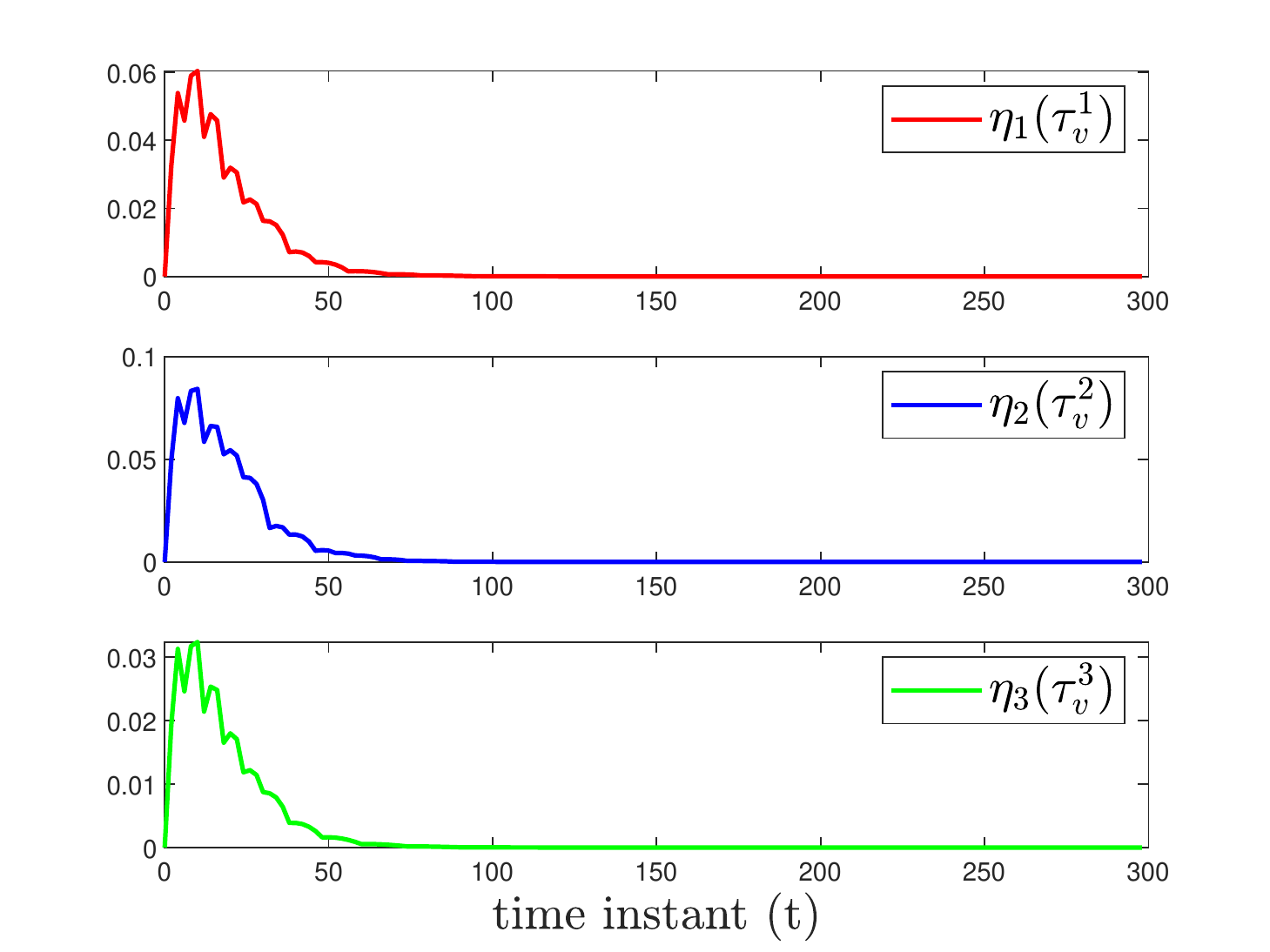}
}	
		\caption{{Trajectories of subsystems $i$ and dynamic variables $\eta_i(\tau_{v}^i)$ under data-driven ETS  \eqref{sys:trigger}.}}
		\label{FIG:x1}
	\end{figure}

\begin{figure}[t]
		\centering
\subfigure{
\includegraphics[scale=0.55]{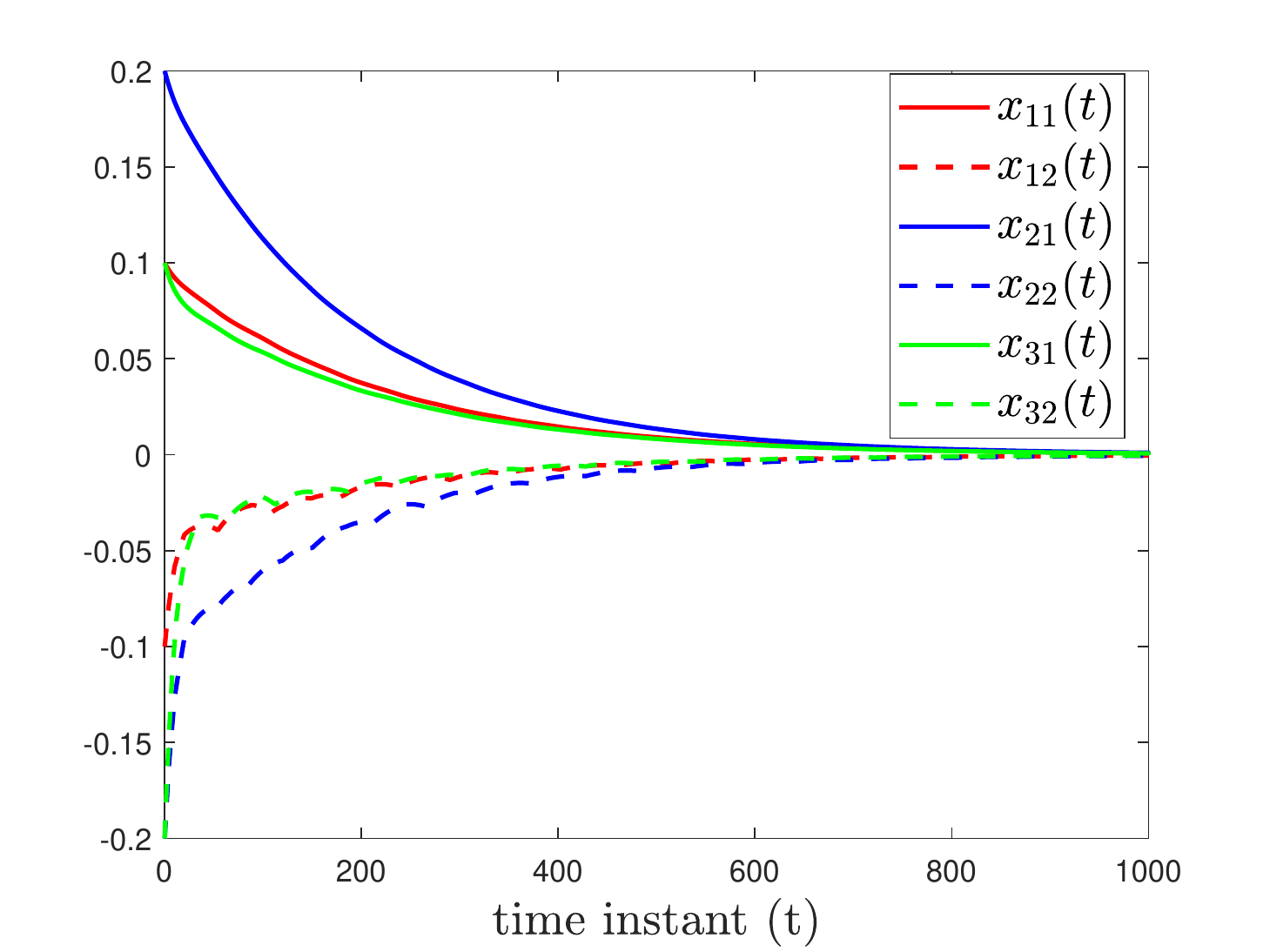}
}
\subfigure{
\includegraphics[scale=0.55]{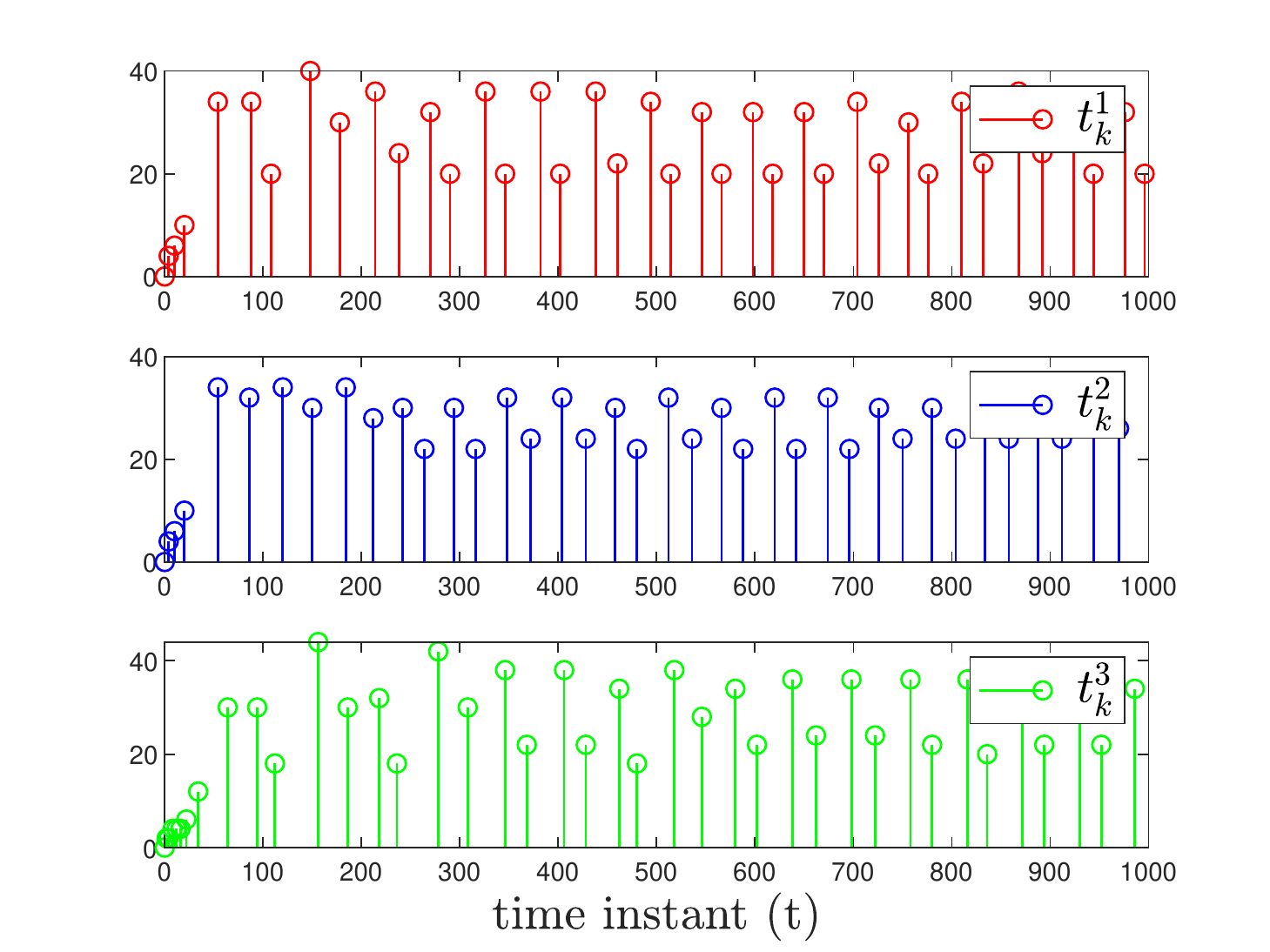}
}	
\subfigure{
\includegraphics[scale=0.55]{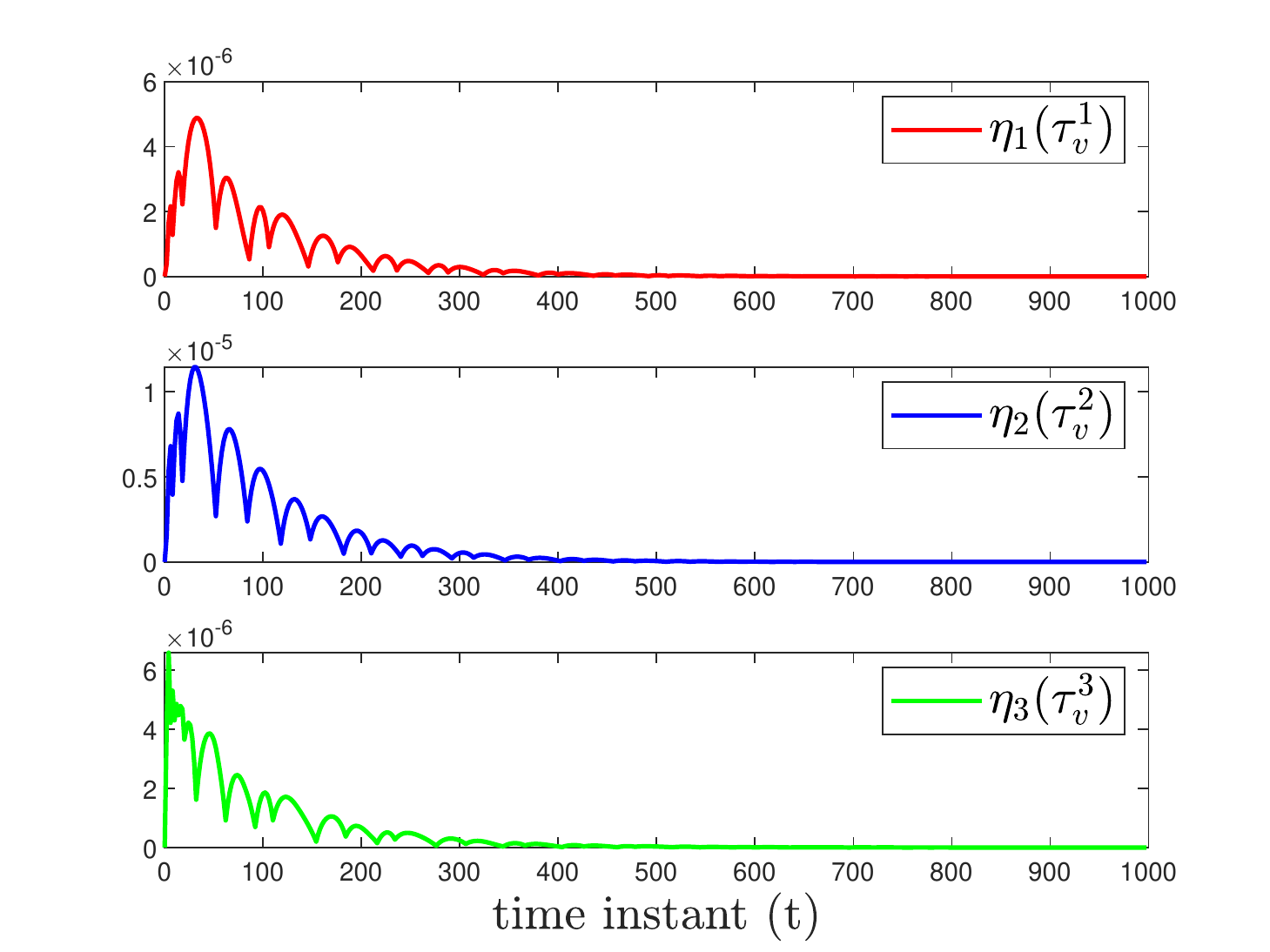}
}	
		\caption{{Trajectories of subsystems $i$ and dynamic variables $\eta_i(\tau_{v}^i)$ under model-based ETS  \eqref{sys:trigger}.}}
		\label{FIG:x1:model}
	\end{figure}

{
\subsection{Testing data-based method}\label{example:data}
}	
In the data-driven controller design, the matrices $A$ and $B$ are supposed to be \emph{unknown}. We set the discretization interval as $T_k=0.01$ and generated $\rho=200$
	measurements $\{x(T)\}^{\rho}_{T=0}$, $\{u(T)\}^{\rho-1}_{T=0}$ from the disturbed system \eqref{sys:data:perturbed}, where the input was generated and sampled randomly from $u(T)\in [-1,~1]$. Besides, the disturbance is assumed to be
distributed randomly over $w(T)\in [-0.001,\,0.001]$, which fulfills Assumption \ref{Ass:disturbance} with $S_d=0$, $Q_d=-I$, and $R_d=0.001^2\rho I$ $(\rho=200)$.
	The matrix $B_w$ was taken as $B_w=0.01I$, which has full column rank.
	Besides, we set the sampling interval $h=0.02$, triggering-related parameters $\sigma_1^i=0.02$, $\sigma_2^{ij}=0.01$ ($j\neq i$), $\theta_i=2$, and $\lambda_i=0.2$ (especially, $\sigma_2^{13}=\sigma_2^{31}=0$).
	Solving the data-based LMIs \eqref{Th:data:LMI1} and \eqref{Th:data:LMI2} in Theorem \ref{Th:data}, 
{the matrices $G_i$}, the controller gains, and triggering matrices were were
computed as follows
\begin{align*}
{G_1}&=G_2=G_3=\left[
	\begin{array}{cccccc}
      0.0096  & -0.0386\\
     -0.0386  &  0.9420\\
      \end{array}
	\right]\\
K_1&=[-327.6018, -77.7775],K_{12}=[-18.3432   ,-0.0282]\\
K_{21}&=[49.5465, -1.2622],~~~~K_2=[-274.3759,  -73.0946]\\
K_{23}&=[4.2697,   -0.4999],~~~~~~K_{32}=[0.0047,   -1.4507]\\
K_3&=[-344.1443, -67.8563],K_{13}=K_{31}=[0, ~0]\\
\Omega_1&=\left[\!\!
	\begin{array}{cccc}
		164.7704 &  17.7939\\
   17.7939  &  5.0145\\
	\end{array}
	\!\!\right],
\Omega_2=\left[\!\!
	\begin{array}{cccc}
		86.1433 &  13.6117\\
   13.6117   & 3.8540\\
	\end{array}
	\!\!\right]\\
\Omega_3&=\left[\!\!
	\begin{array}{cccc}
		121.2177 &  17.9945\\
   17.9945  & 4.4548\\
	\end{array}
	\!\!\right].
\end{align*}
	
The proposed dynamic triggering scheme \eqref{sys:trigger} was numerically tested using the system in \eqref{example:sample:system}, with the initial condition $x_1(0) = [0.1,\, -0.1]^{\top}$, $x_2(0) = [0.2,\, -0.2]^{\top}$, $x_3(0) = [0.1,\, -0.2]^{\top}$
over the time interval $t\in[0,\,300]$.
The simulation results of all subsystems' state trajectories (the top panel), triggered events (the middle panel), along with the dynamic variables $\eta_i(\tau_{v}^i)$ (the bottom panel) are reported in Fig. \ref{FIG:x1}.
	Obviously, both the system states and the dynamic variables converge to zero,
	demonstrating the correctness of the proposed distributed data-driven triggering and control schemes. It should also be pointed out that \emph{only} $34$ measurements for the subsystem $1$, $44$  for the subsystem $2$, and $31$ for the subsystem $3$ were sent to distributed controllers \eqref{feedbackC}, while a total of $150$ data were sampled for each subsystem.
	This validates the effectiveness of the data-driven ETSs in saving communication resources, while achieving distributed control of interconnected subsystems.

\begin{table}[tp]
\caption{{Number of transmitted data under data-driven ETS \eqref{sys:trigger} with different $\theta_i$ over $t \in [0,~300]$.}}
\begin{center}      
\setlength{\tabcolsep}{4pt}
\renewcommand\arraystretch{1.2}
\begin{tabular}{lcccccccccccccccccc}
\hline\noalign{\smallskip}
$\theta_i$ & 2 & 3 &10
&$100$&$200$&$300$&$1000$&2000&$10^4$\\
\hline\noalign{\smallskip}
Subsystem $1$ & 34 & 34 &36  & 39 & 39&40& 40
& 40 &40 \\
Subsystem $2$ & 44 & 45 &51  & 53 &67&74&76
& 76 &76 \\
Subsystem $3$ & 31 & 32 &33   & 37 &38&38&38
& 38 & 38  \\
Total         & 109 & 111 & 120  & 129 &144&152&154 & 154 &154 \\
\hline\noalign{\smallskip}
\end{tabular}
\end{center}
\label{Tab:theta}
\end{table}

\begin{table}[tp]
\caption{{Number of transmitted data under different ETSs over $t \in [0,~300]$.}}
\begin{center}      
\setlength{\tabcolsep}{4pt}
\renewcommand\arraystretch{1.2}
\begin{tabular}{lcccccccccc}
\hline\noalign{\smallskip}
ETSs & Subsystem $1$ & Subsystem $2$ & Subsystem $3$ & Total \\
\hline\noalign{\smallskip}
Data-driven \eqref{sys:trigger}& 34& 44 & 31 &109\\
\eqref{sys:trigger} with $\sigma_2^{ij}=0$& 59 & 40  &  49&148 \\
\eqref{sys:trigger} with $\sigma_1^{i}=0$&44 & 45 &  41& 130 \\
\hline\noalign{\smallskip}
\end{tabular}
\end{center}
\label{Tab:compare}
\end{table}

{
(\emph{Simulations for different $\theta_i$.})
The number of transmitted data under the data-driven ETS \eqref{sys:trigger} with different $\theta_i$ over $t \in [0,~300]$ were listed in Table
\ref{Tab:theta}. As $\theta_i$ increases from $2$ to $10^4$,  the amount of transmissions grows from $109$ to a fixed value $154$. 
The main reason is that our ETS  \eqref{sys:trigger} reduces to
the static decentralized ETS \cite{SHI2019} if $\theta\rightarrow \infty$, which also confirms the statement in Remark \ref{generaN}.
}

\begin{figure}[t]
		\centering
\subfigure{
\includegraphics[scale=0.55]{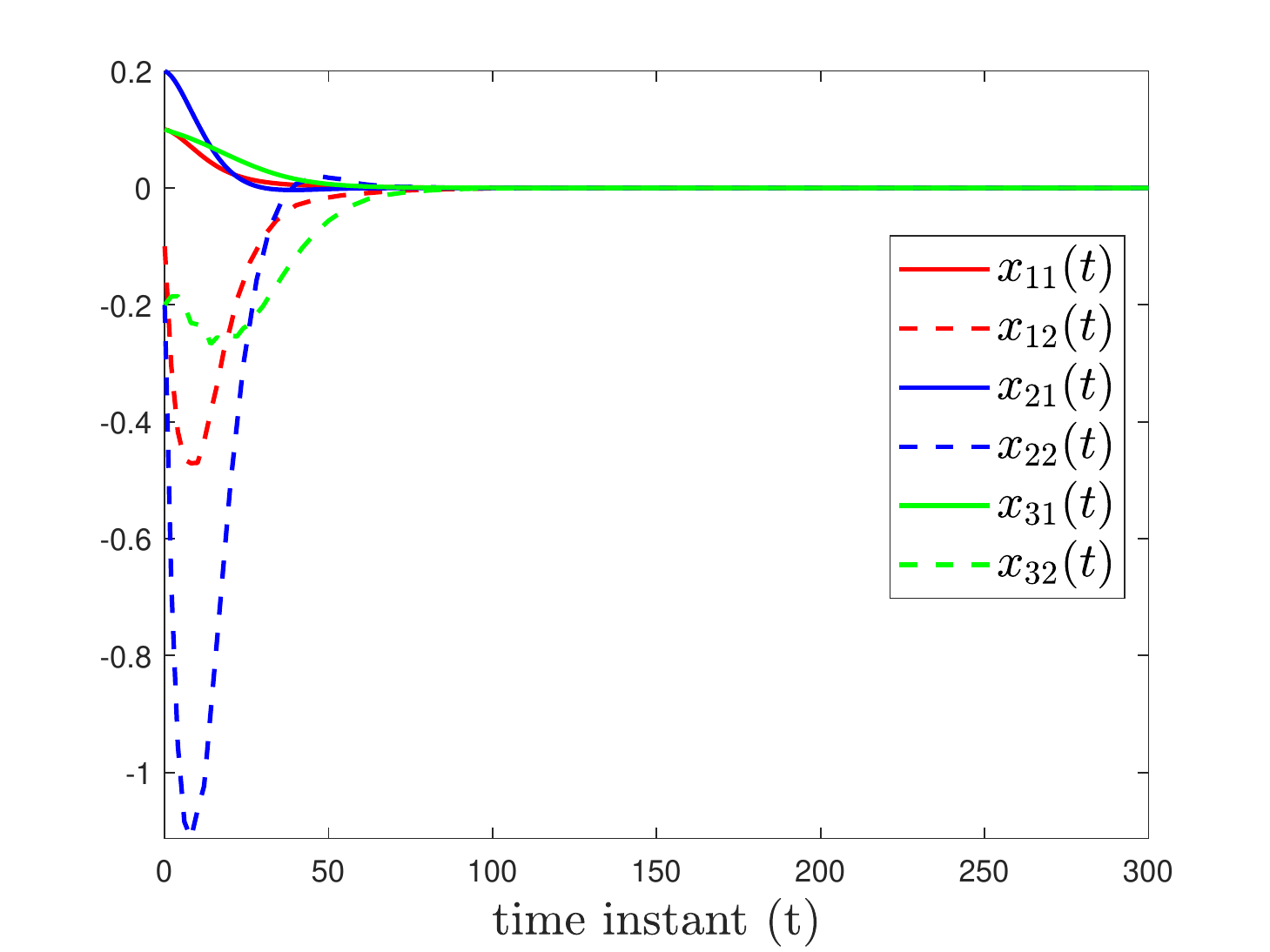}
}
\subfigure{
\includegraphics[scale=0.55]{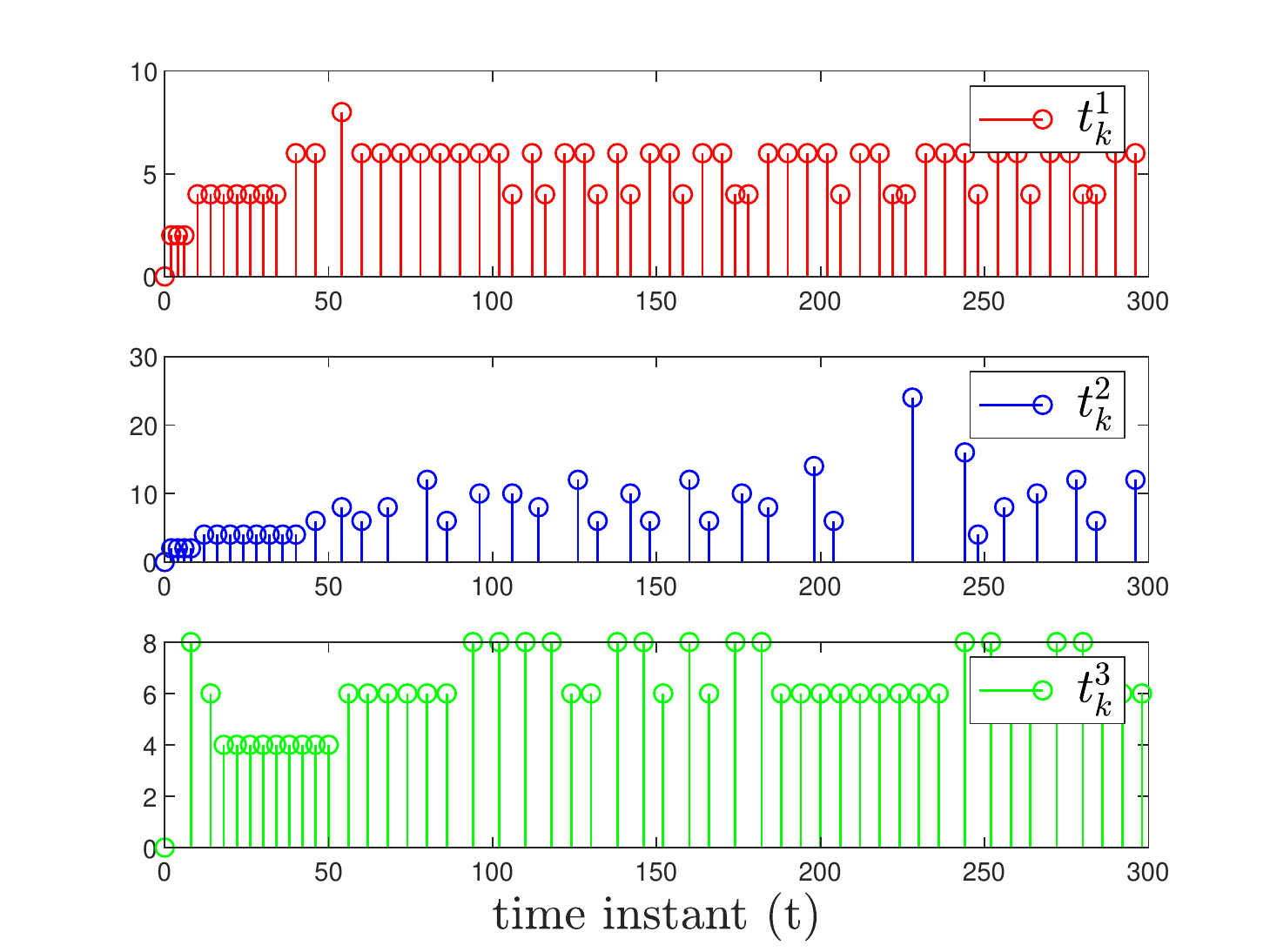}
}	
		\caption{{Trajectories of subsystems $i$ and triggered events under data-driven ETS  \eqref{sys:trigger} with $\sigma_2^{ij}=0$.}}
		\label{FIG:x:data:dec}
	\end{figure}

\begin{figure}[t]
		\centering
\subfigure{
\includegraphics[scale=0.55]{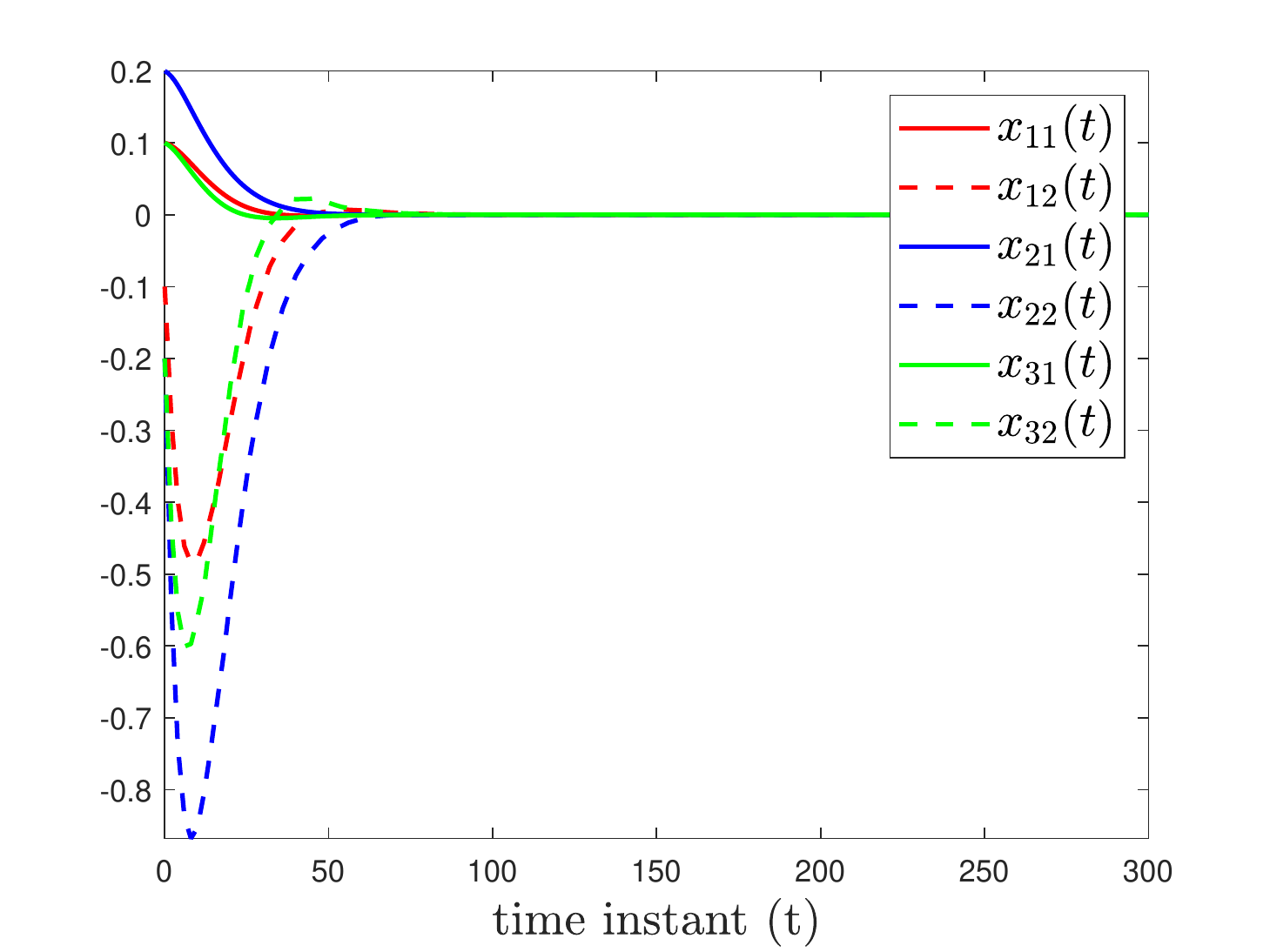}
}
\subfigure{
\includegraphics[scale=0.55]{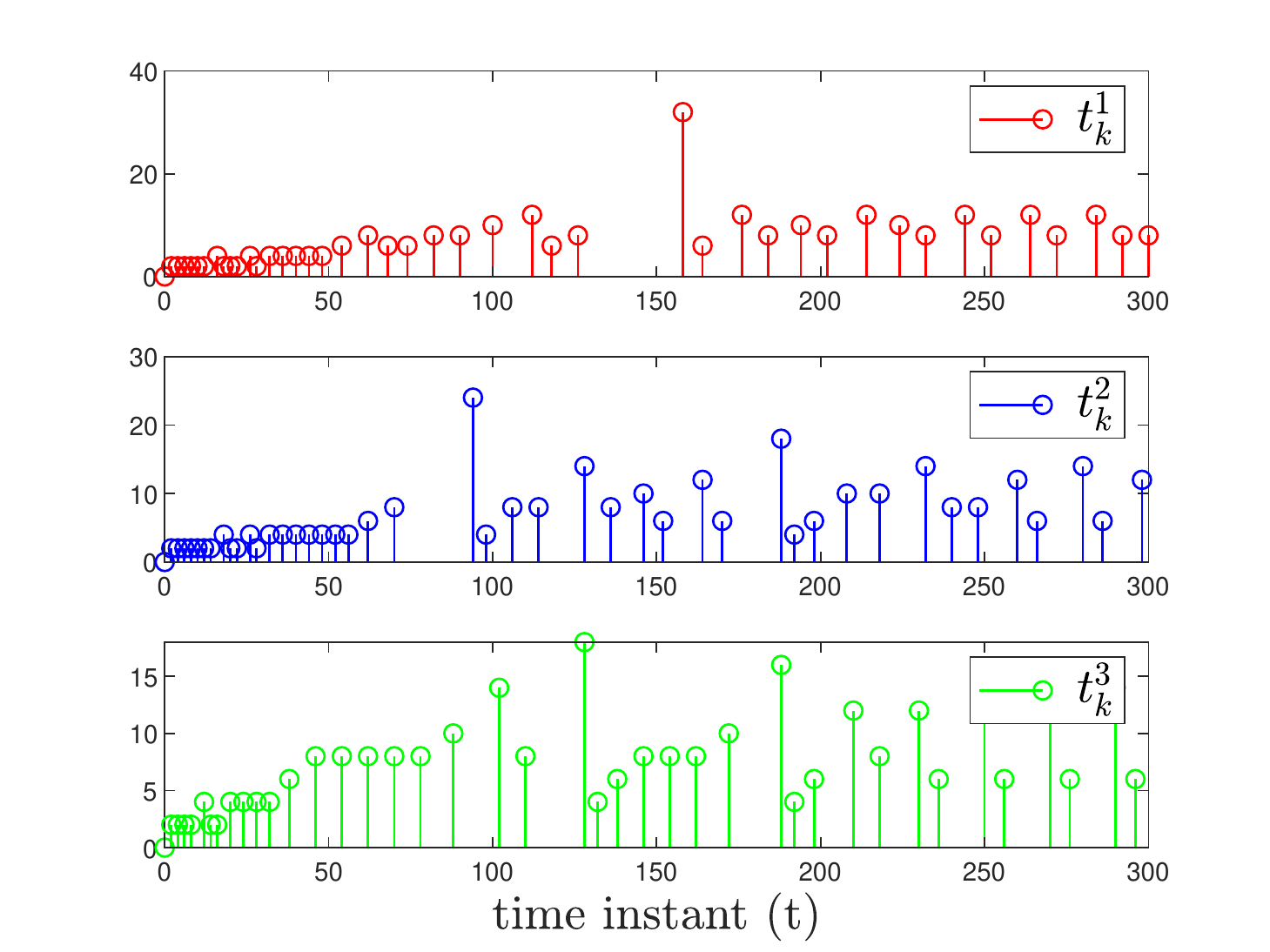}
}	
		\caption{{Trajectories of subsystems $i$ and triggered events under data-driven ETS  \eqref{sys:trigger} with $\sigma_1^{i}=0$.}}
		\label{FIG:x:data:dis}
	\end{figure}
{
\subsection{Testing the model-based method}\label{example:model}

In this case, the matrices $A$ and $B$ of system \eqref{example:sample:system} are assumed \emph{known}. By Theorem \ref{Th:design:model} with the same triggering parameters as in Section \ref{example:data}, the fixed distributed controller gains and triggering matrices were computed as follows
\begin{align*}
K_1&=[-31.8664,  -33.8136],~K_{12}=[-4.3728,   -1.6128]\\
K_{21}&=[-4.4261,   -1.6324],~~~~K_2=[-27.2294,  -32.8590]\\
K_{23}&=[-4.4261,   -1.6324],~~~~K_{32}=[-4.3728,   -1.6128]\\
K_3&=[-31.8664,  -33.8136],~K_{13}=K_{31}=[0, ~0]\\
\Omega_1&=\left[\!\!
	\begin{array}{cccc}
		0.0120  &  0.0094\\
    0.0094  &  0.0121\\
	\end{array}
	\!\!\right],~
\Omega_2=\left[\!\!
	\begin{array}{cccc}
		0.0098  &  0.0084\\
    0.0084  &  0.0117\\
	\end{array}
	\!\!\right]\\
\Omega_3&=\left[\!\!
	\begin{array}{cccc}
		0.0120  &  0.0094\\
    0.0094  &  0.0121\\
	\end{array}
	\!\!\right].
\end{align*}

Then, under the above designed controller and ETS matrices, the trajectories of each subsystem and the extra dynamic variable $\eta_i(\tau_v^i)$ were
depicted in Fig. \ref{FIG:x1:model} with the same initial states in Fig. \ref{FIG:x1}.
Our model-based method (cf. Theorem \ref{Th:design:model}) also ensures that all subsystems and $\eta_i(\tau_v^i)$ converge to zero.
Note in the middle of Fig. \ref{FIG:x1:model} that only $42$ sampled data ($13$, $13$, $16$ for subsystems $1$, $2$, $3$, respectively) were sent to the controllers, which are smaller than
the number in  the middle of Fig. \ref{FIG:x1} (totaling $109$). However, the settling time at the top of Fig. \ref{FIG:x1:model} (around $t=1,000$) is much longer than that in Fig. \ref{FIG:x1} ($t=150$).
The main reason is that Theorem \ref{Th:design:model} has less conservatism than Theorem \ref{Th:data} at the expense of the system performance, which is discussed in Remark \ref{Remark:conservatism}.
}

{
\subsection{Comparing with centralized and distributed ETSs}\label{sec:dec:dis:compare}}

{
This part considers system \eqref{example:sample:system} under the data-driven decentralized ETS (cf. \eqref{sys:trigger} with $\sigma_2^{ij}=0$) and the data-driven distributed ETS (cf. \eqref{sys:trigger} with $\sigma_1^{i}=0$).
Using the same collected state-input data and the triggering parameters as in Section \ref{example:data} except for setting $\sigma_2^{ij}=0$,  we obtained the following controller gains and triggering matrices by Theorem \ref{Th:data} under the decentralized ETS
\begin{align*}
K_1&=[-375.7169,  -76.9418],~K_{12}=[-69.5245,   -1.7845]\\
K_{21}&=[52.7192 ,   2.9506],~~~~~~~K_2=[-561.7601,  -75.7685]\\
K_{23}&=[-65.0679,   -1.8875],~~~K_{32}=[125.9659,    3.1476]\\
K_3&=[-396.4133,  -77.2102],~K_{13}=K_{31}=[0, ~0]\\
\Omega_1&=\left[\!\!
	\begin{array}{cccc}
		275.4520  & 29.3695\\
   29.3695   & 6.4819\\
	\end{array}
	\!\!\right],~
\Omega_2=\left[\!\!
	\begin{array}{cccc}
		417.3350 &  40.3493\\
   40.3493  &  8.2633\\
	\end{array}
	\!\!\right]\\
\Omega_3&=\left[\!\!
	\begin{array}{cccc}
		388.6930  & 32.1356\\
   32.1356 &   7.0680\\
	\end{array}
	\!\!\right].
\end{align*}

Similarly, we co-designed the matrices of the controller and the triggering scheme under the data-driven distributed ETS with
$\sigma_1^{i}=0$ as follows
\begin{align*}
K_1&=[-445.1191,  -72.4965],K_{12}=[-5.4204,    3.8416]\\
K_{21}&=[-25.6825,    2.5192],~~~~~K_2=[-449.6961,  -75.5986]\\
K_{23}&=[62.9174,   -1.3544],~~~~~K_{32}=[-73.8190,   -1.6715]\\
K_3&=[-489.3250,  -75.0031],K_{13}=K_{31}=[0, ~0]\\
\Omega_1&=\left[\!\!
	\begin{array}{cccc}
		126.4117  & 13.4461\\
   13.4461   & 2.5826\\
	\end{array}
	\!\!\right],~
\Omega_2=\left[\!\!
	\begin{array}{cccc}
		116.7566 &  13.4964\\
   13.4964  &  2.4656\\
	\end{array}
	\!\!\right]\\
\Omega_3&=\left[\!\!
	\begin{array}{cccc}
		257.1805 &  14.8794\\
   14.8794  &  3.4479\\
	\end{array}
	\!\!\right].
\end{align*}

Simulating the system from the same initial states in Fig. \ref{FIG:x1} and the triggered events under the data-driven decentralized (with $\sigma_2^{ij}=0$) and under the distributed ETSs (with $\sigma_1^{i}=0$), the states are drawn in Figs. \ref{FIG:x:data:dec} and \ref{FIG:x:data:dis}, respectively. In Table \ref{Tab:compare}, we list
the numbers of transmitted data under the two ETSs as well as \eqref{sys:trigger}.
From the  trajectories in Figs. \ref{FIG:x1}, \ref{FIG:x:data:dec},  \ref{FIG:x:data:dis},  and Table \ref{Tab:compare}, it is obvious that the data-driven ETS \eqref{sys:trigger} generates less transmissions (totaling $109$) than other two mentioned triggering schemes (total $148$ for $\sigma_2^{ij}=0$ and $130$ for $\sigma_1^{i}=0$, respectively), while all steady-states are kept at the same level (where the settling time is around $t=150$). This phenomenon actually tickles the statement in Remark \ref{generaN} by introducing the consensus error $\sum_{j\neq i}^N \sigma_2^{ij} \left[x_i(\tau_v^i)-x_j(\tau_v^i)\right]^\top \Omega_i \left[x_i(\tau_v^i)-x_j(\tau_v^i)\right]$
into the threshold function of \eqref{sys:trigger}, helpful in reducing the transmission frequency compared to the decentralized ETS. }
{
\subsection{Comparison between  centralized and distributed controllers}\label{sec:compare:controller}
}
\begin{figure}[t]
		\centering
\includegraphics[scale=0.55]{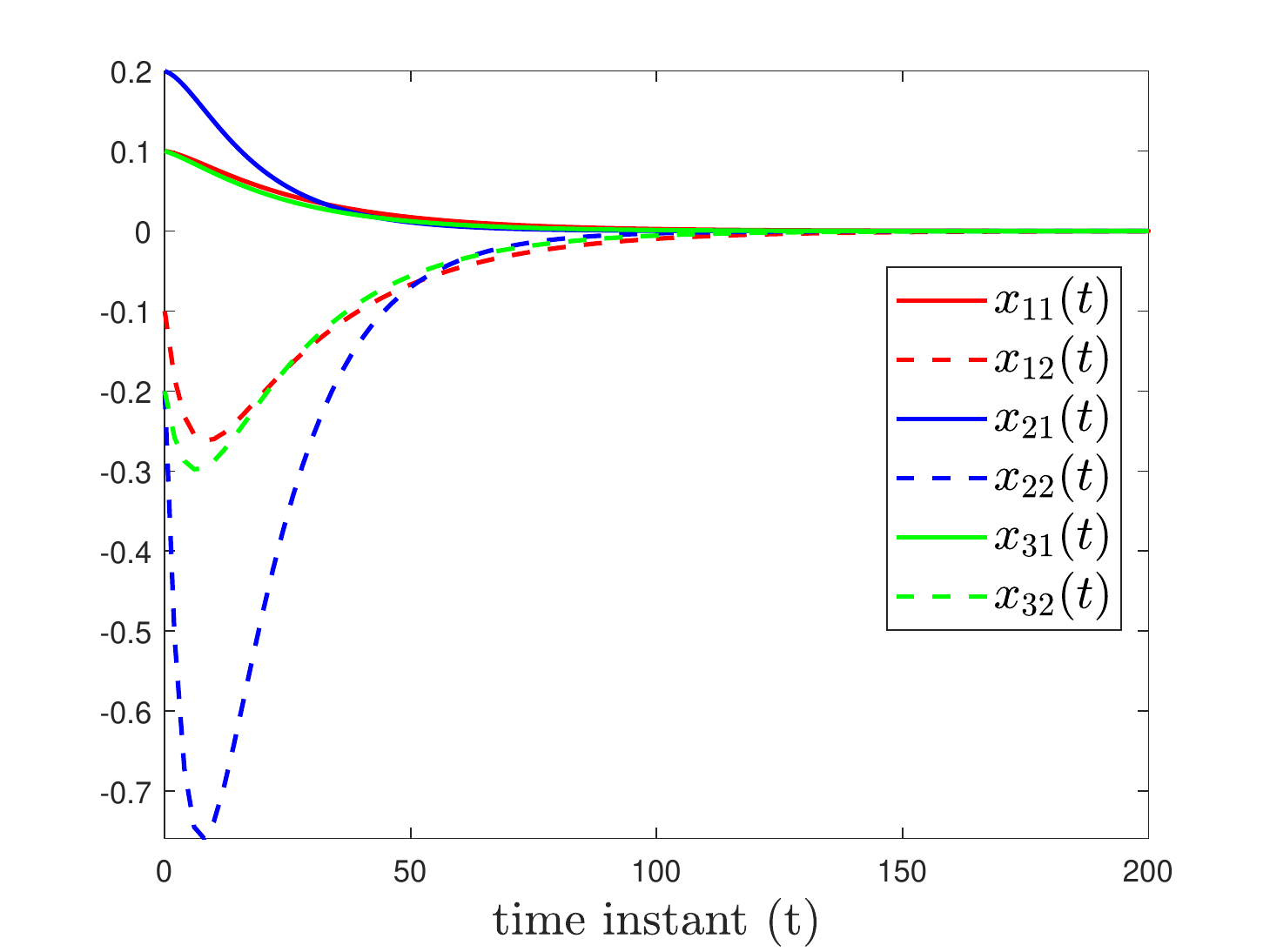}	
		\caption{{Trajectories of subsystems $i$ under decentralized controller \eqref{feedbackC} with $K_{ij}=0$ and periodic transmission scheme.}}
		\label{FIG:x:data:dec:sam}
	\end{figure}

\begin{figure}[t]
		\centering
\includegraphics[scale=0.55]{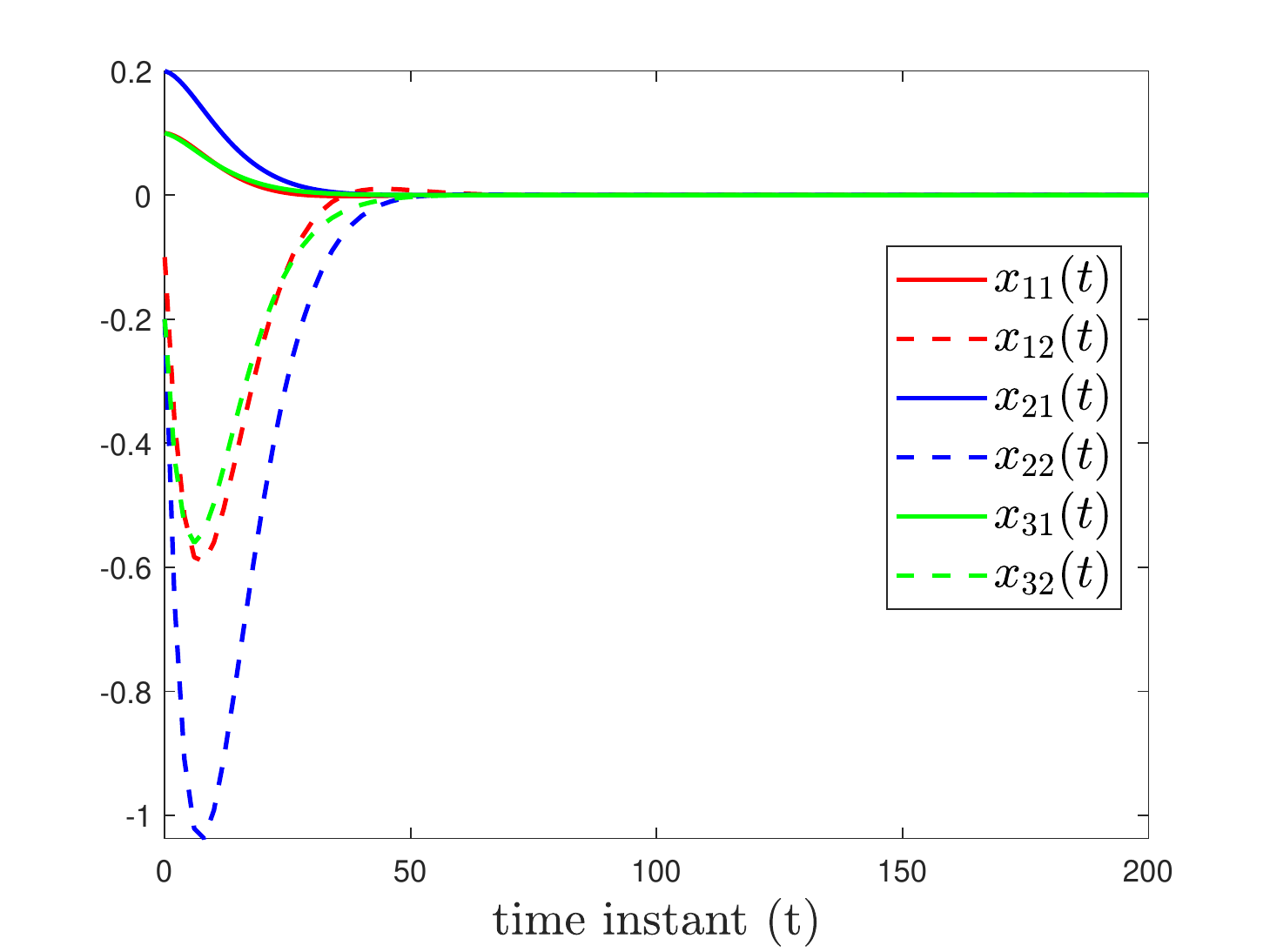}	
		\caption{{Trajectories of subsystems $i$ under distributed controller \eqref{feedbackC} and periodic transmission scheme.}}
		\label{FIG:x:data:dis:sam}
	\end{figure}	

This subsection explores the difference between the data-driven distributed controller \eqref{feedbackC} and the decentralized controller (cf. \eqref{feedbackC} with $K_{ij}=0$) on the
system performance of \eqref{example:sample:system}.
To eliminate the impact of the triggering strategy on the system performance,
we set  $\sigma_2^{ij}=0$ and $\sigma_1^{i}=0$. Therefore, the ETS \eqref{sys:trigger} boils down to the periodic transmission scheme with sampling interval $h=0.02$. Using Theorem \ref{Th:data} and the state-input data of Section \ref{example:data},
we computed the decentralized controller gains as follows
\begin{align*}
K_1&=[-275.8791,  -81.5002]\\
K_2&=[-406.5251,  -84.0427]\\
K_3&=[-296.9997,  -78.1404].
\end{align*}
And, the distributed controller gains were designed  as
\begin{align*}
K_1&=[-674.3840,  -79.6814],~K_{12}=[21.1120,    2.1379]\\
K_{21}&=[-74.5995,  -0.0603],~~K_2=[-573.8955,  -79.0597]\\
K_{23}&=[120.1398,   -0.9242],~~~~K_{32}=[9.6358,   -3.2821 ]\\
K_3&=[-661.8083,  -81.2399],K_{13}=K_{31}=[0, ~0].
\end{align*}

Employing the same initial points in Fig. \ref{FIG:x1},  trajectories of the system \eqref{example:sample:system} under the above decentralized and distributed controllers and the periodic transmission scheme are depicted in Figs. \ref{FIG:x:data:dec:sam} and \ref{FIG:x:data:dis:sam}, respectively. Note that the settling time of system \eqref{example:sample:system} in  Fig.  \ref{FIG:x:data:dis:sam} is around $t=50$, which is
smaller than  $t=150$ in Fig. \ref{FIG:x:data:dec:sam}; namely,
 the distributed controller yields faster convergence rate than the decentralized one. This proves the correctness of the points made in Remark \ref{Remark:controller}.

	\section{Concluding Remarks}\label{sec:conclusion}
	This paper has considered distributed event-triggered control of interconnected discrete-time systems from a data-driven vantage point.  Data-based system representation was developed, based on which a model- and data-driven co-designing approach of the triggering matrix and the controller gain was provided.  Closed-loop system stability under the proposed distributed data-driven event-triggered control scheme was analyzed leveraging a novel looped-functional.
	Finally, a numerical example was provided to corroborate the efficacy of the proposed ETS in saving communication resources, as well as the validity of our co-designing methods.
{Moreover, it was certificated by several numerical comparisons that
the proposed data-driven distributed control strategy has better performance than existing  decentralized and distributed strategies in reducing data transmissions while maintaining desired system performance.}

	\bibliographystyle{IEEEtran}
	\bibliography{cas-refs}

\end{document}